# Photospheric Helium Abundance in Cool Giants – A Comprehensive Study

B. P. Hema[1,*]

and

Gajendra Pandey[1]

[1]*Indian Institute of Astrophysics, Koramangala II Block, Bengaluru, Karnataka-560034, India;*
*hemabp.phy@gmail.com*

## ABSTRACT

High-resolution optical spectra of sixteen red giants, two early asymptotic giant branch (AGB) stars, and two supergiants, having no/minimal to super lithium (Li)-rich abundances, are analyzed to investigate the helium (He)-enhancement. The spectra of eight giants were obtained from the Himalayan Chandra Telescope, and for the rest of the program stars the spectral data were collected from various public archives. Our detailed abundance analyses of the program stars involve the determination of stellar parameters and abundances for about 20 elements among the key abundances of He, Li, C, N, O and the $^{12}$C/$^{13}$C ratios. The difference in the Mg-abundance derived from Mg I lines and the MgH band, and the difference in carbon abundance from C I and the CH-band, are used as a clue to the mild hydrogen-deficiency/helium-enhancement. From this analysis, four red giants, an early-AGB star and a supergiant star were found to be enhanced in helium. All these He-enhanced stars are also found to be super Li-rich except for the supergiant. Since the He-rich red giants are Li-rich as well, this implies, that He-enrichment is accompanied by Li-enrichment but not vice-versa. This is the first spectroscopic measurement of photospheric He-abundance in normal and Li-rich field giants. The Li-enrichment is observed across the giant branch from RGB-bump (KIC 9821622) to AGB phase, unlike that expected from the RGB-tip to RC-phase. A plausible scenario for the enrichment of He and as well as Li in giants is the fresh synthesis of Li in the interiors of giants and dredging up along with He to the surface, from deeper layers. However, there could be multiple scenarios operating in tandem. This analysis of He- and Li-enrichment along with other key elements provides more insights to decipher the mystery of Li-enrichment in giants.

## 1. Introduction

Lithium (Li) is one of the key elements in the study of stellar evolution that its observed abundance provides a clue to the star's evolutionary stage (Iben 1967). Li is a fragile element that can get destroyed at a mere 2.5 million Kelvin. The Li abundance in the atmospheres of main-sequence (MS) stars like the Sun, comes from its primordial material and gradually gets diluted due to the mixing process. As the star evolves to the giant branch, the Li abundance is reduced by about 10 to 100 times its primordial content. There are a handful of giants that are exceptions to this theory; these giants have a Li abundance of about 1000 times more than that observed in the Sun. The observed abundance of Li in most of the giants is about, $\log \epsilon(\mathrm{Li}) < 1.5$ dex but in Li-rich giants it ranges between $1.5 < \log \epsilon(\mathrm{Li}) < 4.0$ dex (Kumar & Reddy 2009; Charbonnel & Balachandran 2000; Reddy & Lambert 2005; Gratton & D'Antona 1989; Lambert et al. 1980; Drake et al. 2002; Luck & Challener 1995; Singh et al. 2020; Brown et al. 1989; Gonçalves et al. 2020; Cayrel de Strobel et al. 1997; Wallerstein & Sneden 1982; Kumar et al. 2011, 2020; Luck & Heiter 2007). The Li-rich stars are mostly low-mass giants with masses $\leq 2\mathrm{M}\odot$ with the metallicity ([Fe/H]) ranging from $-0.5 < [\mathrm{Fe/H}] < +0.2$. These Li-rich giants are found all across the giant branch of post-main sequence (MS) evolution, that is from the base of the red giant branch (RGB), the RGB-Bump, the RGB, the tip of the RGB, the red clump (RC) phase, and also AGB stars.

The source of Li-enrichment has been explored for decades. This is yet another study exploring Li-enrichment in giants with a different approach. One such scenario is the fresh production of Li in the giants through the Cameron-Fowler reaction (Cameron & Fowler 1971) and dredging up to the surface along with the other H-burnt materials such as helium, nitrogen (N) and $^{13}C$. In this study, we are exploring the He-enhancement in the program stars, to investigate if the He- and Li-enrichment is correlated.

The He-enhancement in the cool giants is investigated using our novel and validated method of measuring the accurate He/H ratio from the magnesium (Mg) features; that is, the Mg I and the subordinate lines of the (0,0) MgH band. Along with the Mg-features, the C-features; the C I lines and the CH-band were also explored for the determination of He/H ratio. Our sample stars (a random sample) include giants with a range in the abundance of Li, from No-Li to Super Li-rich abundances from different evolutionary stages of RGB, and two are from the AGB phase. The analyses of the program stars are based on the high-resolution optical spectra. By conducting detailed abundance analyses, the elemental abundances for the standard He/H ratio (He/H=0.1) and also for enhanced He/H ratio are derived. The key elements include He, Li, C, N, O and the $^{12}\mathrm{C}/^{13}\mathrm{C}$ ratio. The methodology, analysis and results are discussed in the following sections. A recent study by Mallick et al. (2025) has explored the correlation between high Li-abundances and strong chromospheric He I $\lambda 10830$Å absorption line strengths in Kepler field giant stars, which we discuss in light of our results.

## 2. Sample selection, Observations and Data reduction

We aim to investigate the He-enrichment in giants with and without Li-enrichment in different stages of their RGB evolution. In total, twenty stars are analyzed in this study. The high-resolution optical spectra for 8 program stars are observed from the Himalayan Chandra Telescope (HCT),

Hanle. Out of eight giants, four are from the Kepler survey[1] for which the evolutionary states are known by asteroseismology (Borucki et al. 2010) and the other four are known local giants. For the remaining twelve program stars, the high-resolution spectra were retrieved from the archives of the ELODIE, ESO: UVIT and FEROS, McDonald Observatory and the Kitt Peak Observatory. The observations and the data handling processes are discussed below. The details of the observations are summarized in Table 1.

### 2.0.1.   The Himalayan Chandra Telescope

High-resolution optical spectra for eight of our program stars were obtained from the Himalayan Chandra Telescope (HCT), Hanle. The 2m HCT is mounted with the high-resolution Hanle Echelle Spectrograph (HESP) (Sriram et al. 2018). HESP is a fibre-fed, cross-dispersed spectrograph that uses a $4K\times4K$ CCD covering the complete visible wavelength range from 3700-10000Å providing a resolving power (R=$\lambda/\Delta\lambda$) of 30000 as well as 60000. The observations with R=30000 were carried out between April 2022 to June 2023 (see Table 1). The data reduction and processing were carried out using the IRAF software package[2]. The ThAr lamp and Quartz blue lamps were used for wavelength calibration and flat fielding, respectively. The high-resolution spectra for the program stars; KIC 9821622, KIC 5184199, KIC 2305930, KIC,12645107, HD 40827, HD 107484, HD 233517, and Ups02 Cas were obtained from the HCT.

### 2.0.2.   Elodie Spectrograph – OHP

ELODIE is a fibre-fed cross-dispersed echelle spectrograph installed at the Observatoire de Haute-Provence 1.93m reflector which has been in operation since 1993. Light from the Cassegrain focus is fed into the spectrograph through a pair of optical fibres. The spectra cover about 3000Å ranging from 3850-6800Å with a resolving power of ∼42000 is recorded on a $1K\times1K$ CCD. The instrument is entirely computer-controlled and a standard data reduction pipeline automatically processes the data (Moultaka et al. 2004). The spectra of the program stars taken from ELODIE archive[3] are: HD 30834, HD 39853, HD 112127, HD 205349, HD 214995 and HD 181475.

---

[1] https://archive.stsci.edu/missions-and-data/kepler

[2] The IRAF software is distributed by the National Optical Astronomy Observatories under contract with the National Science Foundation (Tody 1986, 1993)

[3] http://atlas.obs-hp.fr/elodie/

### 2.0.3. The UVES and FEROS – ESO

UVES, the Ultraviolet and Visual Echelle Spectrograph is a high-resolution optical spectrograph of the very large Telescope installed at the Unit Telescope-2 (UT2) of 4 UTs of 8.2m diameter of ESO. It is a dual beam, cross-dispersed echelle spectrograph with resolutions of up to R=80,000 to 115,000. The wavelength coverage of about 1000Å in the blue arm and about 2000 or 4000Å in the red arm, with the wavelengths between 3200Å to 10500Å (Dekker et al. 2000). The reduced spectra of the program stars: HR 334 and HR 4813 were retrieved from the ESO-UVES archives[4].

FEROS, the Fiber-fed Extended Range Optical Spectrograph (Kaufer et al. 1999) is a high-resolution spectrograph installed at the MPG/ESO 2.2m telescope. It is a fiber-fed visible spectrograph that operates in Cassegrain mode of focus covering the wavelengths ranging from 3500Å to 9200Å at a resolving power, R=48,000. The reduced spectrum for the program stars HR 6766 and HD 19745 were retrieved from the ESO-FEROS archives.

### 2.0.4. McDonald Observatory

The high-resolution optical spectrum of the super Li-rich cool giant HD 77361 was obtained from the McDonald Observatory. The spectrum obtained with the 2.7 m Harlan J. Smith Telescope and Tull coude cross-dispersed echelle spectrograph (Tull et al. 1995) is at a resolving power of about R $(\lambda/\Delta\lambda) = 50000$. The spectrum was reduced using the IRAF software package following the standard procedure.

### 2.0.5. Kitt Peak observatory

For our standard-giant star Arcturus, the spectrum was obtained from the Archives of Nicholas U. Mayall 4-meter Telescope at Kitt Peak National Observatory (KPNO). These observations were made with the Coude Feed Telescope on Kitt Peak with the spectrograph in the Echelle mode at a resolving power of about R$(\lambda/\Delta\lambda) = 150000$ with the wavelength coverage of from 3727 to 9300Å (Hinkle et al. 2000).

---

[4]https://archive.eso.org/scienceportal/home

Table 1.   The program stars with their log of observations.

| Star | RA | Dec | $V_{mag}$ | Obs./Instr. | Date of Obs. | S/N ratio | R=$\lambda/\Delta\lambda$ |
|---|---|---|---|---|---|---|---|
| KIC 9821622 | 19 08 36.15 | +46 41 21.25 | 12.2 | HCT, Hanle | 10 June 2022 | 200 | 30000 |
| KIC 5184199 | 19 23 48.42 | +40 20 36.18 | 10.1 | HCT, Hanle | 9 April 2023 | 200 | 30000 |
| KIC 2305930 | 19 28 25.63 | +37 41 23.24 | 11.0 | HCT, Hanle | 9 April 2023 | 150 | 30000 |
| KIC 12645107 | 19 17 12.5 | +51 45 11.38 | 11.4 | HCT, Hanle | 10 June 2022 | 130 | 30000 |
| HD 40827 | 06 05 08.23 | +59 23 35.56 | 6.3 | HCT, Hanle | 9 April 2023 | 300 | 30000 |
| HD 107484 | 12 21 10.08 | +41 28 58.90 | 7.7 | HCT Hanle | 9 April 2023 | 250 | 30000 |
| HD 233517 | 08 22 46.71 | +53 04 49.23 | 9.7 | HCT Hanle | 9 April 2023 | 200 | 30000 |
| Ups02 Cas | 00 56 39.90 | +59 10 51.80 | 4.6 | HCT Hanle | 17 September 2022 | 250 | 30000 |
| Arcturus | 14 15 39.67 | +19 10 56.67 | -0.05 | Mayall Tel. and FTS | 1993-94 | 1000 | 150000 |
| 1HD 77361 | 09 01 11.41 | -26 39 49.37 | 6.2 | MacDonald Obs. | 25 Dec 2009 | 300 | 50000 |
| 1HD 214995[a] | 22 41 57.45 | +14 30 59.01 | 5.9 | OHP/ELODIE | 10 July 2000 | 200 | 42000 |
| HD 39853[b] | 05 54 43.62 | -11 46 27.08 | 5.6 | OHP-ELODIE | 03 Oct. 1996 | 140 | 42000 |
| HD 112127[c] | 12 53 55.74 | +26 46 47.98 | 6.9 | OHP-ELODIE | 27 May 2003 | 280 | 42000 |
| HD 205349[d] | 21 33 17.88 | +45 51 14.45 | 6.2 | OHP-ELODIE | 3 Oct 2003 | 150 | 42000 |
| HD 30834[e] | 04 52 25 | +36 41 53 | 4.8 | OHP-ELODIE | 05 Oct. 1996 | 180 | 42000 |
| HD 181475[f] | 19 20 48.31 | -04 30 09.0 | 7.0 | OHP-ELODIE | 02 July 2001 | 170 | 42000 |
| HR 334 (HD 81797) | 01 08 35.39 | -10 10 56.15 | 3.45 | ESO-VLT-U2, UVES | 10 March 2016 | 272 | 115000 |
| HR 4813 (HD 110014) | 12 39 14.76 | -07 59 44.04 | 4.6 | ESO-VLT-U2,UVES | 10 March 2016 | 356 | 107200 |
| HR 6766 (HD 165634) | 18 08 04.97 | -28 27 25.53 | 4.5 | ESO-FEROS | 11 May 2012 | 300 | 48000 |
| HD 19745 | 03 07 03.24 | -65 26 56.80 | 9.1 | ESO-FEROS | 7 Dec 2016 | 110 | 48000 |

[1]Notes.  Observer : Program No. - (a) Guillout: elodie 20000709/0010, (b) Soubiran: elodie 19961003/00178, (c) Liang: elodie 20030527/0013, (d) Lebre: elodie 20031003/0023, (e) Soubiran: elodie 19961005/00138, (f) Catala: elodie 20010702/0018

### 3. Abundance Analyses

The abundance analyses for all the program stars were conducted using their high-resolution optical spectra obtained as discussed above. For conducting the abundance analyses, the equivalent widths for weak and moderately strong lines of several species that are clean and not severely blended for both neutral and ionized states were measured. The very strong lines were avoided as they are saturated and do not fall in the linear part of the curve of growth. The line list was adopted from Hema et al. (2018); Hema & Pandey (2020); Hema et al. (2020) for which the log $gf$ values were validated using the Solar and the Arcturus spectrum. The new lines included in this work are given in Table A.1, in Appendices. The atomic data for these new lines were adopted from the NIST atomic and molecular database[5] and from Reddy et al. (2003). The atomic data adopted were also validated by deriving the abundances for Arcturus.

For the determination of stellar parameters and the elemental abundances the LTE line analysis and spectrum synthesis code MOOG (Sneden 1973) and the ATLAS9 (Kurucz 1998) plane parallel, line-blanketed LTE model atmospheres were used. The microturbulence ($\xi_t$) is derived using Fe I lines having a similar excitation potential and a range in equivalent width, weak to strong, giving the same abundance. The effective temperature ($T_{eff}$) is determined using the excitation balance of Fe I lines having a range in lower excitation potential (LEP). The $T_{eff}$ and $\xi_t$ were fixed iteratively. The process was carried out until both returned zero slope for the derived Fe abundances. By adopting the determined $T_{eff}$ and $\xi_t$, the surface gravity (log $g$) is derived. The log $g$ is fixed by demanding the same abundances from the lines of different ionization states of a species, known as the ionization balance. The log $g$ is derived mainly from the lines of Fe I/Fe II, Ti I/Ti II, and Sc I/Sc II. As the abundances of Sc I/Sc II are not available for all the stars, the emphasis is mainly given to Fe I/Fe II and Ti I/Ti II. Then, the mean log $g$ was adopted. By adopting these stellar parameters, the abundances were derived for about 23 elements including the key elements C, N, O and $^{12}C/^{13}C$ ratios using both the atomic and molecular lines. The details of the line list and the methodology for deriving the C, N, O abundances and the $^{12}C/^{13}C$ ratios are discusssed in Appendix A. In Figure A.1, the top panel shows the synthesis of the CN-band for deriving N-abundance, and the bottom left panel for deriving $^{12}C/^{13}C$.

Uncertainties on the $T_{eff}$ and $\xi_t$ are estimated by changing the $T_{eff}$ in steps of 25 K and $\xi_t$ in steps of 0.05 km s$^{-1}$, for the adopted stellar parameters ($T_{eff}$, log $g$ and $\xi_t$). The change in $T_{eff}$ and $\xi_t$ and the corresponding deviations in abundance, from the zero slope abundance, of about $1\sigma$ error, is obtained. This change is adopted as the uncertainty on these parameters. The adopted $\Delta T_{eff} = \pm 50$ K and $\Delta \xi_t = \pm 0.1$ km s$^{-1}$ (see Figure 4 of Hema & Pandey (2020), section 3). The uncertainties on log $g$ is the standard deviation from the mean value of the log $g$ determined from different species, which is about $\pm 0.1$ (cgs units). The uncertainty adopted on the abundances are the standard deviation of the line-to-line scatter. The uncertainty on the abundances due to the uncertainties on the stellar parameters are $< 0.05$ dex (Hema et al. 2018). Since, the root-mean-square (RMS) of the errors due to the stellar parameters is less than the line-to-line scatter, the later were adopted.

---

[5]https://physics.nist.gov

The non-LTE (NLTE) correction, $\Delta_{NLTE}$=log$\epsilon_{NLTE}$-log$\epsilon_{LTE}$, ranges from 0.03 to 0.09 dex for various Fe I lines considering electron and H I collisions, and the correction is about 0.01 for Fe II (Mashonkina et al. 2011). The Fe I with the excitation energy of the lower level ($E_{LEP}$) $< 1$ eV, the $\Delta_{NLTE}$ is from 0.04 to 0.09 dex, while $\Delta_{NLTE}$=0.04 dex is the upper limit for the Fe I lines with $E_{LEP} > 1$eV. Our line list includes the Fe I lines with the $E_{LEP} >$1eV. For Fe I the NLTE effects on the abundances are expected to be smaller for stars with solar-type metallicities, than the metal-deficient stars (Bergemann et al. 2012). For this non-LTE correction of Fe I lines, the downward revision of the star's adopted $T_{\rm eff}$ by 50K restores the ionization balance for Fe I/Fe II. This variation in the Fe I and Fe II abundances could also arise due to the uncertainties in the log $gf$ values. However, the adopted uncertainty on the derived Fe abundance for the program stars is the standard deviation of the line-to-line scatter of abundances ($0.07 < \Delta$ log$\epsilon$(Fe I) $<$ 0.16 dex, except 0.05 dex for $\alpha$ Boo). This adopted uncertainty is the upper limit and accounts for that due to the stellar parameters and the non-LTE corrections. The $\Delta_{NLTE}$ for the Ti I/Ti II abundance is within ≈0.02 dex (Mallinson et al. 2022). For the Sc lines, the $\Delta_{NLTE}$ is about $-0.06$ (Mashonkina 2024). The non-LTE corrections are significant in very metal-poor dwarfs and giants with [Fe/H]$< -2.0$ dex. The positive corrections of Fe species and the negative corrections of Sc, remove the bias, and our determined $T_{\rm eff}$ and log $g$ are accurate within the uncertainties.

The log g values were also derived by the parallax method as well as asteroseismology to support our spectroscopic determination. For the distances derived from the GAIA parallaxes, the log $g$ values were derived using the standard relation. These log $g$ values agree within 0.1 dex with those derived by us spectroscopically. The log $g$ values were also derived from the seismic parameters; the frequency of maximum oscillation power ($\nu_{max}$) and the large frequency separation ($\Delta\nu$), with the standard relation. For 3 of the 4 Kepler field objects of our program stars, the log $g$ values derived by asteroseismology are in excellent agreement with those derived spectroscopically. Hence, the log $g$ values derived from both these methods are in good agreement with the spectroscopic determinations within the adopted uncertainty limits.

## 4. Spectral analysis of MgH and CH Bands

### 4.1. The MgH Band

The He-enhancement in the program stars is investigated using the procedure developed by Hema & Pandey (2014); Hema et al. (2018, 2020). The Mg abundance derived from the Mg I lines and from the subordinate lines of the (0, 0) MgH band from the observed stellar spectrum is expected to be same within the uncertainties. In some of our program stars, there is a difference of 0.3 dex or more in the derived Mg-abundances from these feature. If this difference is not reconcilable by the uncertainties on the stellar parameters, then it is attributed to the mild
H-deficiency/He-enhancement in the stellar atmosphere.

Spectra of the program stars are synthesized from 5175–5179Å which includes the unblended
subordinate lines of the (0,0) MgH band (see Figures 1 for example). For spectrum synthesis, the
atomic lines were compiled from the standard atomic data sources along with those identified by
Hinkle et al. (2000). The $(0,0)$ MgH molecular line list was adopted from Hinkle et al. (2013).
Synthetic spectra were generated by combining the LTE spectral line analysis/synthesis code
MOOG (Sneden 1973), and the ATLAS9 Kurucz (1998) plane parallel, line-blanketed LTE model
atmospheres with convective overshoot. The spectrum of Arcturus, a typical red giant, was
synthesized to validate the adopted gf–values of the atomic/molecular lines. Using the spectrum
synthesis code, *synth* in MOOG, the high-resolution optical spectrum of Arcturus was
synthesized for the adopted stellar parameters and the abundances derived by us that agree with
those given by Ramírez & Allende Prieto (2011). The synthesized spectrum was convolved with a
Gaussian profile with a width that represents the broadening due to macroturbulence and the
instrumental profile. Minimal adjustments were made to the abundances of the atomic lines to
obtain the best fit to the observed high-resolution optical spectrum of Arcturus (Hinkle et al.
2000). The changes in the log gf–values were not more than 0.1 dex. A reasonably good fit was
obtained for the MgH molecular lines for the adopted isotopic values from McWilliam & Lambert
(1988). A good match of the synthesized Arcturus spectrum to the observed high–resolution
spectra validates the adopted line list for the adopted stellar parameters of Arcturus. These
checks on the published analysis of Arcturus are taken as evidence that our implementation of the
code MOOG, the LTE models, and the adopted line list were successful for the syntheses of the
red giants' spectra. Hence, the spectrum of a program star was synthesized following the above
procedure. An Mg I line at 5711Å in the spectra of our program stars is also synthesized to
support the Mg abundance derived from the Mg I lines by equivalent width analysis (refer to the
bottom right panel of Figure A.1, in Appendix).

For a program star, the Mg abundance derived from the Mg I lines and from the subordinate lines
of the MgH band were compared. If the Mg abundance derived by MgH subordinate lines is <0.3
dex than that derived from the Mg I lines, this difference can be reconciled by changing the stellar
parameters within the uncertainties (see Table 2 of Hema & Pandey (2020)). And, for some of the
program stars, the Mg abundance derived from the MgH bands is less by 0.3 dex or more than
that derived from the Mg I lines. This difference ($\geq 0.3$ dex) cannot be reconciled by synthesizing
the MgH bands by making the changes to the stellar parameters within the uncertainties and also
for a slightly higher margin on the uncertainty of the $T_{\mathrm{eff}}$. The Mg abundance derived from the
MgH band with a change of about $\pm 100$ K on the adopted $T_{\mathrm{eff}}$, and the uncertainty on the
adopted log $g$ of $\pm 0.1$, also does not provide a comparable Mg abundance from that of the Mg I
lines. This difference ($\geq 0.3$ dex) in Mg abundance derived from Mg I lines and from the MgH
band is unacceptable. Hence, this difference is attributed to the lower H- or He-enhancement in
the atmosphere of the program star. Using the models constructed for differing He/H ratios, the analyses were carried. The He/H ratios are derived by enforcing the fact that the derived Mg abundances from the Mg I lines and from the subordinate lines of the MgH band must be the same for the adopted model atmosphere and stellar parameters.

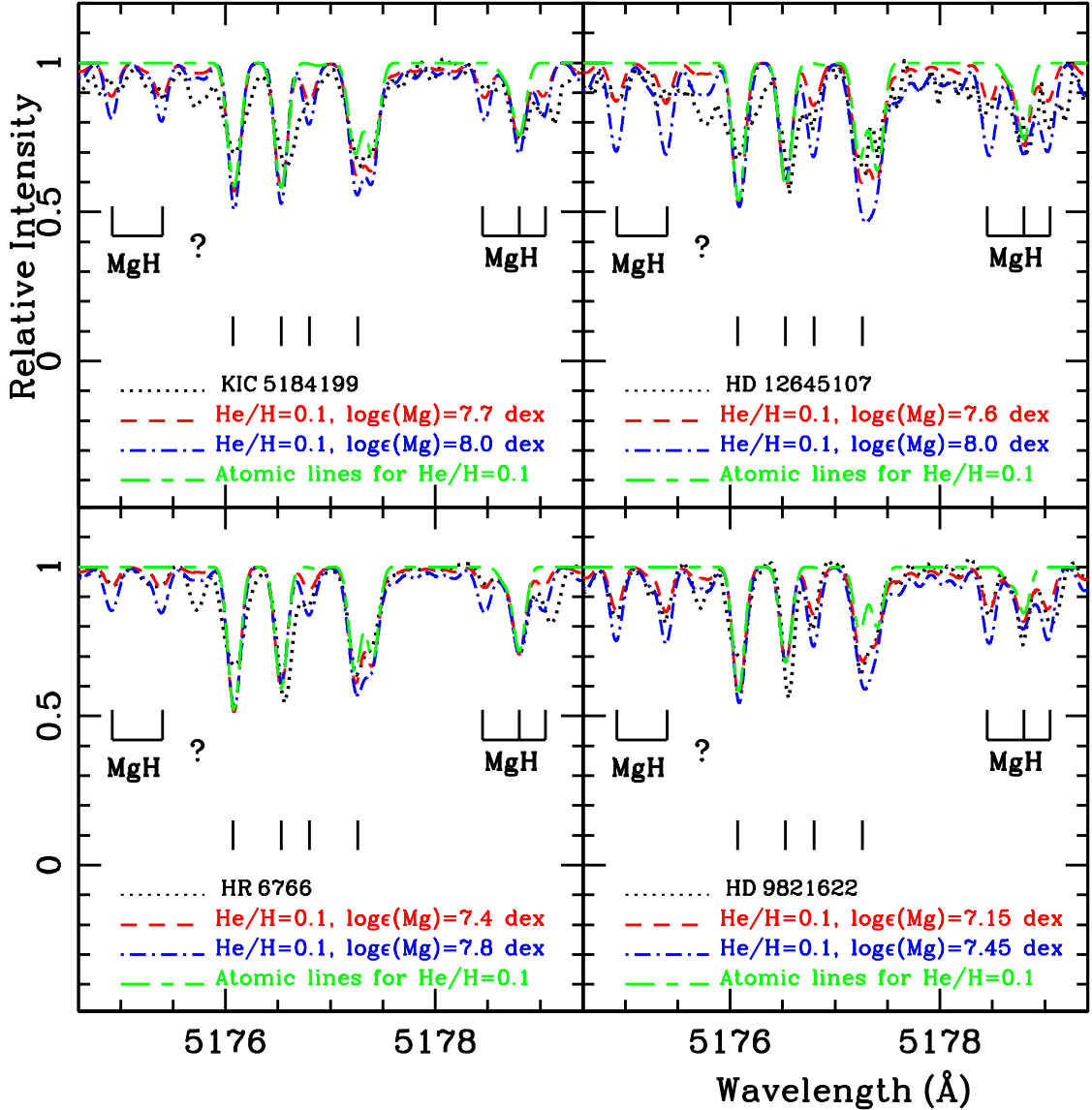

Fig. 1.— Figure show the observed and the synthetic spectra of MgH band. The dotted black line is the observed spectrum of the program star. The best fit synthesized spectrum is shown with dashed red line for the stars' derived He/H ratio (=0.1) and the corresponding Mg-abundance (from Mg I and MgH). The synthesis is also shown with dash dotted blue line for slightly higher Mg-abundance that is not providing the fit. The synthesis for pure atomic lines is shown with the long and short dashed green line. The MgH lines are marked. The vertical lines are the atomic lines, Co I, Ni I, V I and Fe I from blue to redward in the spectrum and the question mark is an unknown line.

As discussed in Section 3, the NLTE correction to the Fe I lines increases the mean Fe abundance by $\approx 0.1$ dex. For this change in Fe I abundance, reducing the $T_{\text{eff}}$ by about 50K from adopted is required to restore the ionization balance. With this downward revision of $T_{\text{eff}}$, the strength of the synthesized MgH band increases for the Mg abundance derived from the Mg I lines. To match the strength of the synthesized MgH band with the observed, the adopted Mg abundance from Mg I lines has to be decreased further. Hence, increasing the difference in the Mg abundance derived from Mg I lines and the MgH band, than that for the adopted stellar parameters from the LTE analysis and making the stars more He-enhanced. The determination of the He/H ratio is discussed in detail in the following section.

## 4.2. The CH-band

In support of our analyses of He-enhancement using the Mg-features, the C-abundance from C I lines and the subordinate lines of the CH(G)-band are also explored. Similar to the analyses of Mg-features, the analyses of the CH-band is carried out by fixing the C-abundance accurately from the C I and [C I] lines.

The CH $X^2\pi$–$A^2\delta$ "G-band" has strong absorption in the wavelength window $\lambda4200\text{-}4400\text{Å}$ with the band head at 4300Å. We selected three wavelength regions that have relatively less contamination of atomic lines: $\lambda4300\text{-}4308\text{Å}$ $\lambda4308\text{-}4315\text{Å}$ and $\lambda4322\text{-}4327\text{Å}$. The carbon abundance determined from 4308-4315Å is given more weight as it is less contaminated by atomic lines relative to the other two regions (Afşar et al. 2012). For our analysis, the $^{12}$CH and $^{13}$CH line lists were kindly provided by Prof. Bertrand Plez (through private communications). Using this line list along with the appropriate $ATLAS9$ model atmospheres and MOOG $synth$ code the spectra were synthesized and matched with the observed spectra. The Arcturus spectrum was synthesized to validate the line list and also the selected spectral regions/spectral lines for a well-determined C-abundance (refer to Figures 2 and 3). For Arcturus, the C-abundance derived from the synthesis of the CH-band and the C I and [C I] lines are in excellent agreement. Using these methods, the spectra of program stars were analyzed for He-enhancement. Since the subordinate lines of $^{12}$CH and $^{13}$CH are blended, and due to the lack of pure $^{13}$CH lines, the $^{12}$C/$^{13}$C ratio is not estimated from this band.

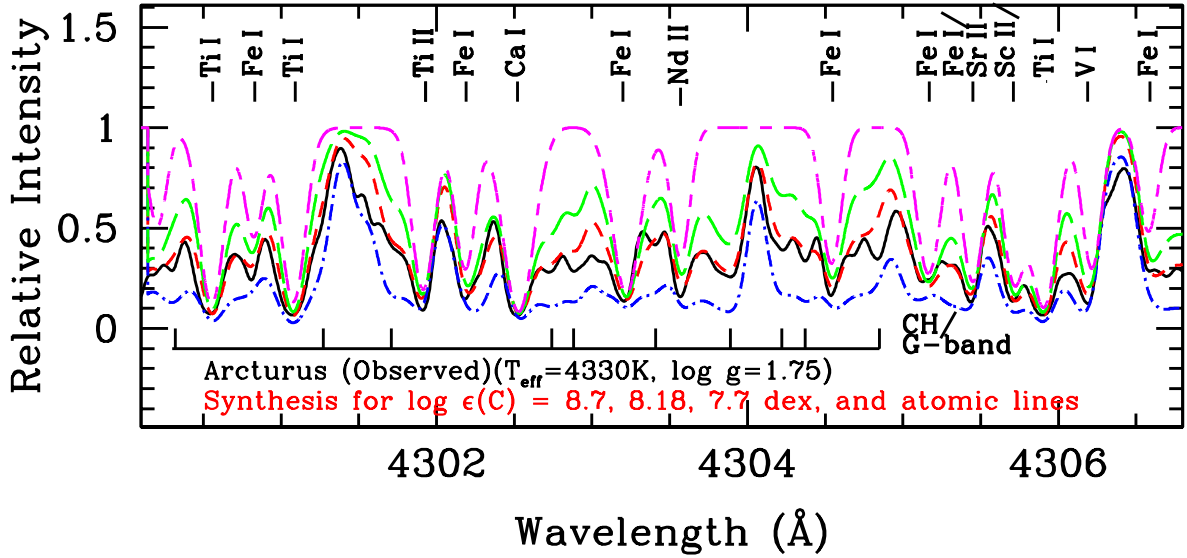

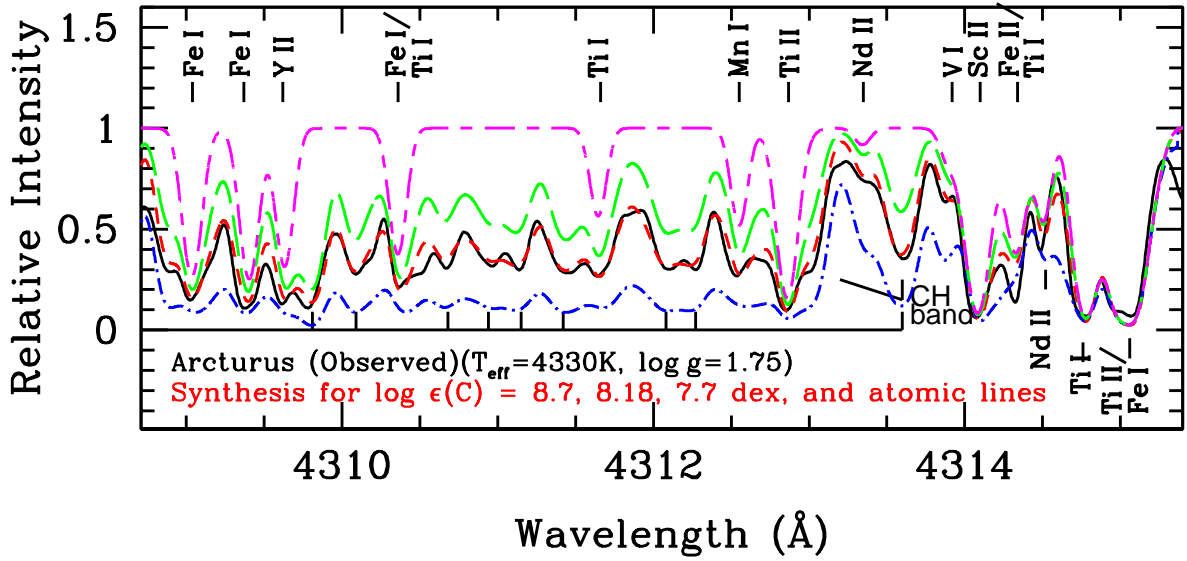

Fig. 2.— Figure shows the synthesis of CH-band for Arcturus in the two wavelength windows: 4300Å to 4307Å (top) and 4308Å to 4315Å (bottom) in comparison with the observed spectrum. The synthesis is shown for three different values of carbon abundances. The synthesis shown are for the best fit (short-dashed line in red) and the other two synthesis shown with long dashed line in green and dash-dotted line in blue. The magenta line with long- and short-dashed line shows the synthesis for pure atomic lines. The atomic lines and the molecular band is marked.

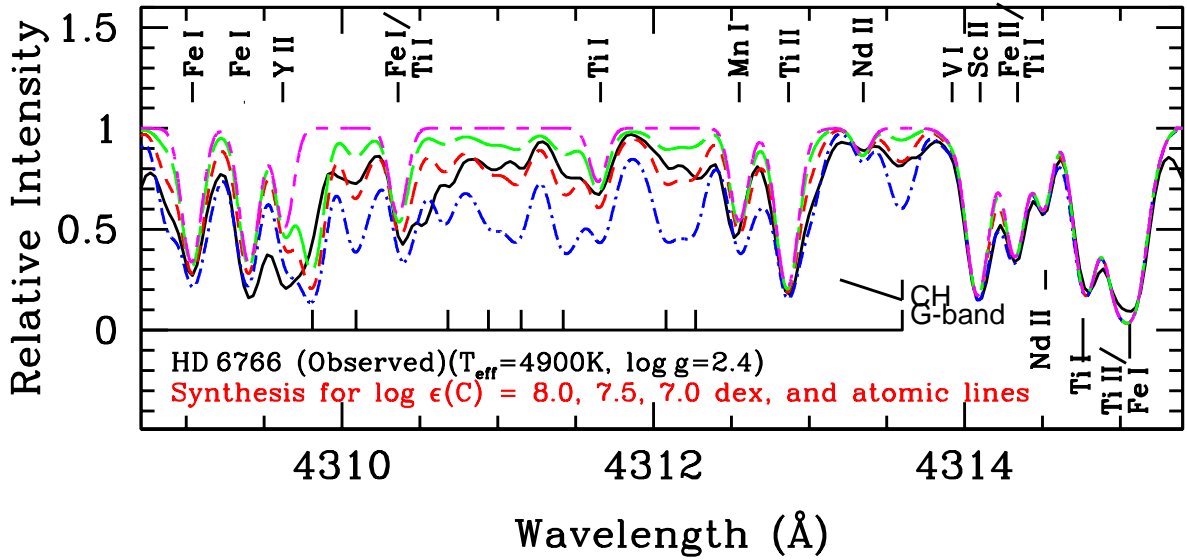

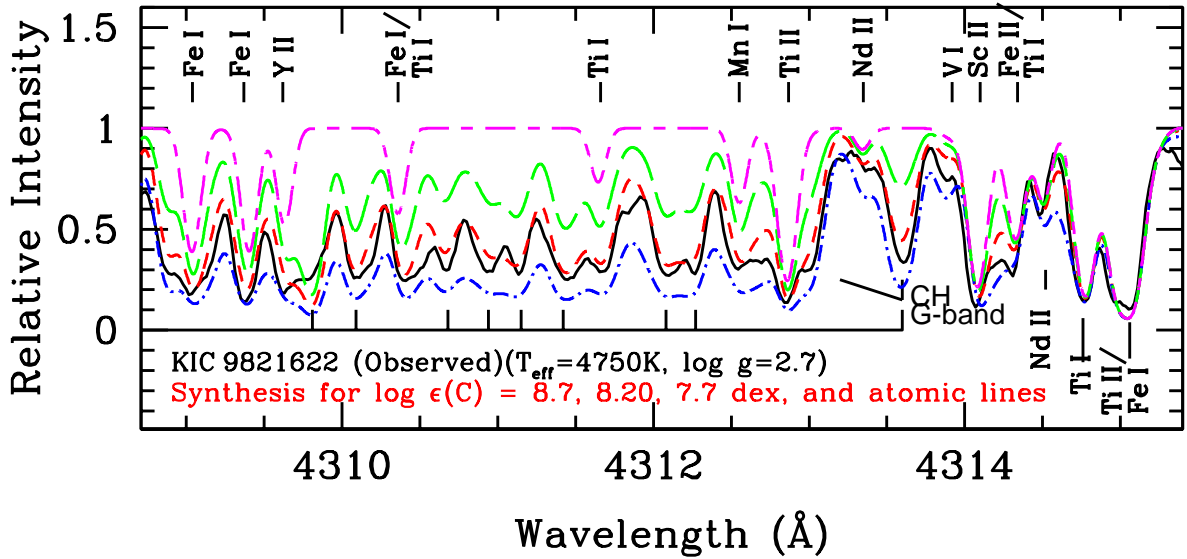

Fig. 3.— Figure shows the synthesis of CH-band in the wavelength window 4307Å to 4316Å for the program stars HD 6766 (top panel) and KIC 9821622 (bottom panel) in comparison with the observed spectrum. The synthesis is shown for three different values of carbon abundances. The synthesis shown are for the best fit (short-dashed line in red) and the other two synthesis shown with long dashed line in green and dash-dotted line in blue. The magenta line with long- and short-dashed line shows the synthesis for pure atomic lines. The atomic lines and the molecular band is marked.

## 5. Determination of He/H ratio

The abundance analyses of the program stars discussed in Section 3 are for normal He/H ratio[6] of 0.1. For deriving the He/H ratio for the program stars the stellar parameters were rederived using a grid of model atmospheres with He/H = 0.15, 0.20, 0.25, 0.30, 0.4, 0.5. The scripts were wrote to iterate and generate Kurucz LTE plane-parallel model atmospheres with a fine grid in He/H ratios from the coarse grid of model atmospheres (Please refer Hema et al. (2020) for details of the model atmospheres and the scripts used).

The input abundances of H and He provided to MOOG, which adopts a model atmosphere computed for a normal He/H ratio of 0.1, are log$\epsilon$(H)=12, and log$\epsilon$(He)=11, respectively. Nevertheless, the input abundances of H and He provided to MOOG are log$\epsilon$(H)=11.894 and log$\epsilon$(He)=11.195, respectively, if a model atmosphere of He/H ratio 0.2 is adopted for analysis.

These abundances of H and He for different He/H ratios are calculated by using the standard normalization relation:

$$\sum_i \mu_i E_i = \mu_H H + \mu_{He} He + \sum_{i=3} \mu_i E_i = 10^{12.15}$$

where $\mu_i E_i$ represents the total mass of an element E, with atomic number $i$, with $\mu_i$ and $E_i$ denoting the atomic mass and abundance by number for the element E, respectively. Assuming that H and He are the most abundant elements in the stellar atmosphere, while all other elements present are in trace amounts, *i.e.*,

$$\sum_{i=3} \mu_i E_i \to 0$$

then, $H + 4He = 10^{12.15}$, since $\mu_H = 1$, and $\mu_{He} = 4$. Note that, conventionally, the log of $H$ is log $\epsilon$(H) and the log of $He$ is log $\epsilon$(He), or, in general log of $E$ is log $\epsilon$(E).

The rederived stellar parameters are not sensitive and are almost independent of the adopted grid's He/H ratio. For the adopted stellar parameters of the program stars, the MgH and the CH-band regions were synthesized for different He/H ratios. The Mg- and C-abundances used for the above syntheses are those derived from the measured equivalent widths of the weak Mg I and C I lines for the adopted model's He/H ratio. Hence, the best fit to the MgH band and the CH band to the observed spectrum determines the adopted model's He/H ratio. Finally, the elemental abundances are derived for the adopted stellar parameters: $T_{eff}$, log $g$, $\xi_t$, and the He/H ratio. Note that the adopted metallicity of the model atmosphere comes from the Fe-abundances derived from the measured equivalent widths of the Fe lines for the star in question, and is an iterative process. The results of our analyses with the adoption of $\alpha$-enhanced model atmospheres having [O/Fe] = 0.5, are in excellent agreement with that of solar-scaled model atmospheres having [O/Fe] = 0.

---

[6] $n_{He}/n_H$

The elemental abundances derived by adopting the model atmospheres with enhanced He (or He/H>0.1) are lower than that derived for the normal He/H ratio of 0.1. Here, decreasing the H-abundance or increasing the He-abundance, that is the He/H > 0.1, lowers the continuous opacity per gram (Sumangala Rao et al. 2011; Hema et al. 2020). Hence, for the same observed strength of the spectral line, the elemental abundance has to decrease (refer to Table 4 for example). In Table 4, the abundances derived for the normal He/H ratio of 0.1, and for 0.4 are given for KIC 2305930. The decrease in abundance is proportional to the amount of H-deficiency or He-enrichment applied in the sense of He/H ratio, except for a few, such as carbon. However, the abundance ratios remain unchanged for most of the elements with few exceptions.

Compared to the MgH band, the CH band is less sensitive to the change in He/H ratio. Therefore, the He/H ratio derived from the MgH band is given more weight. Figure 4 illustrates the variation in Mg abundance from Mg I lines and the MgH band, and C abundance from C I and the CH band with respect to the He/H ratio for the program stars. Once the stellar parameters and metallicity are determined and fixed, the strength of the MgH band primarily depends on the Mg abundance and the He/H ratio. Consequently, the uncertainty in the He/H ratio mainly arises from the uncertainty on the Mg abundance.

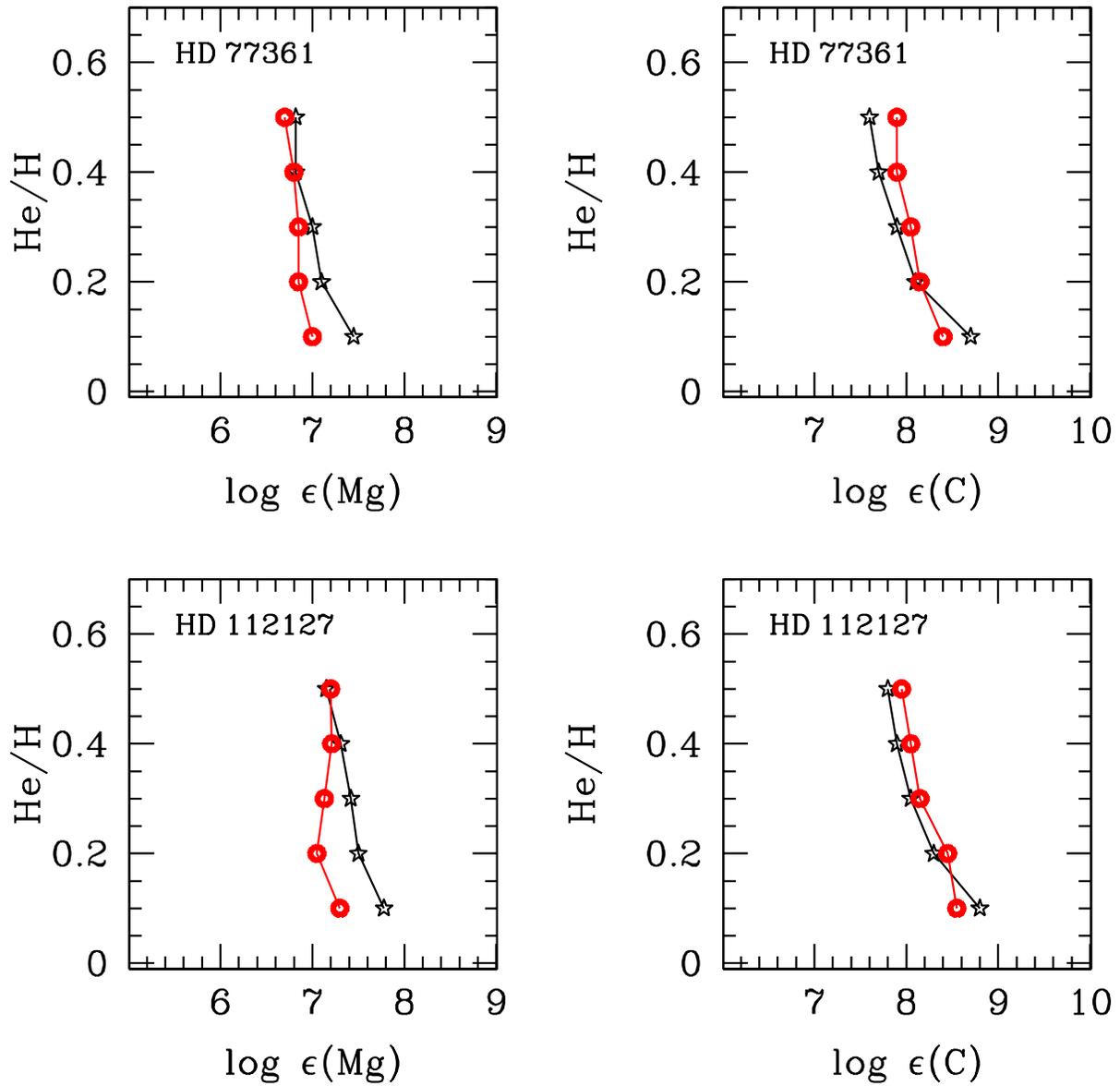

Fig. 4.— Figure shows the change in abundance of Mg and C with the He/H ratios. The top panel is shown for the program star HD 77361 and the bottom panel for HD 112127. The left-hand side panels show the change in the Mg abundance derived from the Mg I lines (black stars) and from the MgH-band (red dots) for different He/H ratios for these program stars. And the right hand panels show the change in C-abundance derived from the C I lines (black stars) and from the CH-band (red dots) for different He/H ratios for these program stars.

Table 2. The program stars in the order of their $T_{\text{eff}}$ with their stellar parameters, $\log \epsilon(\text{Li})$, He/H ratio and the $v\sin i$ are given.

| Star | $T_{\text{eff}}$ (K) | $\log$ g (cm s$^{-2}$) | $\log$ L/L$_\odot$ | M/M$_\odot$ | $\log \epsilon$ (Li) (dex) | He/H | $v\sin i$ (km s$^{-1}$) |
|---|---|---|---|---|---|---|---|
| KIC 5184199 | 5000±50 | 2.7±0.1 | 1.59 | 2.16 | < +0.5 | 0.1 | $\cdots$ |
| KIC 12645107 | 4925±50 | 2.7±0.1 | 1.76 | 1.05±0.04 | 3.72 | 0.1 | 1.5±0.04[b] |
| HR 6766 | 4900±50 | 2.4±0.1 | 2.2 | 3.6 | 1.18 | 0.1 | <5[c] |
| KIC 9821622 | 4750±50 | 2.7±0.1 | 1.6 | 1.46 | 1.54 | 0.1 | 1.01[d] |
| Ups02 Cas | 4750±50 | 2.5±0.1 | 1.74 | $\cdots$ | < −0.3 | 0.1 | $\cdots$ |
| KIC 2305930 | 4725±50 | 2.5±0.1 | 1.59 | 0.92±0.11 | 4.03 | 0.4 | 13.09[e] |
| HD 19745 | 4650±50 | 2.5±0.1 | 1.65 | 2.6[a] | 4.55 | 0.1 | 1.0[f] |
| HD 40827 | 4620±50 | 2.6±0.1 | 1.80 | 2.5±0.3 | 1.87 | 0.1 | 1.8[g] |
| HD 214995 | 4600±50 | 2.5±0.1 | 1.58 | $\cdots$ | 3.27 | 0.1 | 4.7±0.3[h] |
| HDE 233517 | 4600±50 | 2.7±0.1 | 2.18 | 1.7±0.2 | 4.24 | 0.1 | 17.6[i] |
| HD 107484 | 4600±50 | 2.5±0.1 | 1.65 | 0.0 | 2.10 | 0.4 | $\cdots$ |
| HR 334 | 4500±50 | 2.3±0.1 | 1.84 | $\cdots$ | 1.37 | 0.1 | 3.5±0.5[j] |
| HR 4813 | 4400±50 | 2.0±0.1 | 2.30 | $\cdots$ | < −0.2 | 0.1 | 2.52±0.4[k] |
| HD 112127 | 4340±50 | 2.15±0.1 | 1.57 | 1.1±0.2 | 3.40 | 0.4 | 1.6[l] |
| $\alpha$ Boo | 4330±50 | 1.75±0.1 | 2.23 | 1.1 | < −1.0 | 0.1 | 3.3[m] |
| HD 205349 | 4330±50 | 1.7±0.1 | 3.63 | >5 | 1.92 | 0.1 | $\cdots$ |
| HD 77361 | 4270±50 | 2.0±0.1 | 1.84 | 1.5±0.2 | 3.90 | 0.4 | $\cdots$ |
| HD 30834 | 4200±50 | 1.82±0.1 | 2.78 | 4.5±0.3 | 2.3 | 0.3 | 2.7[g] |
| HD 181475 | 4100±50 | 2.3±0.1 | 3.68 | $\cdots$ | −0.8 | 0.5 | 8.4[g] |
| HD 39853 | 4000±50 | 1.2±0.1 | 2.77 | 1.5±0.3 | 3.90 | 0.1 | 3.1[g] |

[1]**Note:** (a) da Silva et al. (1995), (b) Kumar et al. (2018), (c) Palacios et al. (2016), (d) Singh et al. (2020), (e) Ceillier et al. (2017), (f) de Medeiros et al. (1996), (g) de Medeiros & Mayor (1999), (h) Frasca et al. (2018), (i) Balachandran et al. (2000), (j) Soto et al. (2021), (k) Jofré et al. (2015), (l) Cayrel de Strobel et al. (1997), (m) Borisov et al. (2023)

## 6.  Program Stars

The program stars were analyzed to explore the He-enhancement and its connection to Li-enrichment, if any. In this section, He-normal and the He-enhanced giants are discussed in the order of their decreasing $T_{\mathrm{eff}}$.

### 6.1.  He-normal giants in the order of decreasing $T_{\mathrm{eff}}$

**KIC 5184199**

KIC 5184199 is a horizontal branch or RC star for which the evolutionary state is known by asteroseismic studies of the Kepler survey (Kepler Mission Team 2009). The Li I line is weak and only yields an abundance upper limit, $\log\epsilon$ (Li) < 0.5 dex upon synthesis. The derived stellar parameters (Table 2) and elemental abundances (Table 3) agree well with the APOGEE spectral analysis (Jönsson et al. 2020; Abdurro'uf et al. 2022). The derived C, N and $^{12}$C/$^{13}$C ratio, along with no detection of Li as are expected for giants. The Mg-abundance from Mg I and MgH-band (see Figure 1), and the C-abundance from C I and CH-band are in excellent agreement for the He/H=0.1 with no enhancement in He (Table 10).

**KIC 12645107**

KIC 12645107 is an RC-star as derived from the Kepler survey. The derived stellar parameters (Table 2) and elemental abundances (Table 3) agree well with those reported by Kumar et al. (2018). The derived Li-abundance is about $\log\epsilon$(Li)=3.72±0.1 dex. The star's C, N and the $^{12}$C/$^{13}$C ratio are as expected for first dredge-up mixing, except for Li. There is no enhancement of He (Figure 1) nevertheless, Li is enhanced.

Table 3. Elemental abundances for program stars in the order of their decreasing $T_{\mathrm{eff}}$. The numbers given in parentheses with the abundances are the number of spectral lines of that species used for analysis.

| Elements | Sun $\log \epsilon(\mathrm{X})_\odot$ | KIC 5184199 $\log \epsilon(\mathrm{X})_{\mathrm{He/H}=0.1}$ | [X/Fe] | KIC 12645107 $\log \epsilon(\mathrm{X})_{\mathrm{He/H}=0.1}$ | [X/Fe] | HR 6766 $\log \epsilon(\mathrm{X})_{\mathrm{He/H}=0.1}$ | [X/Fe] | KIC 9821622 $\log \epsilon(\mathrm{X})_{\mathrm{He/H}=0.1}$ | [X/Fe] |
|---|---|---|---|---|---|---|---|---|---|
| H | 12.00 | 12.00 | $\cdots$ | 12.00 | $\cdots$ | 12.00 | $\cdots$ | 12.00 | $\cdots$ |
| He | 10.93 | 11.00 | $\cdots$ | 11.00 | $\cdots$ | 11.00 | $\cdots$ | 11.00 | $\cdots$ |
| Li | 1.05 | <0.50(1) | −0.50 | 3.72±0.1(2) | 2.77 | 1.18(1) | 0.18 | 1.54(1) | 0.99 |
| C | 8.43 | 8.27±0.04(4) | −0.11 | 7.88±0.07(3) | -0.45 | 7.45(1) | −0.93 | 8.15±0.05(4) | 0.22 |
| CH | 8.46 | 8.40±0.1(3) | −0.01 | 7.88±0.15(3) | -0.45 | >7.0(3) | $\cdots$ | 8.20±0.03(3) | 0.24 |
| $^{12}$C/$^{13}$C[a] | 89 | 7±2 | $\cdots$ | 6±2 | $\cdots$ | >10 | $\cdots$ | 6±1 | $\cdots$ |
| N | 7.83 | 8.0±0.1(2) | 0.22 | 8.30(CN) | 0.57 | 8.67±0.09(7) | 0.89 | 8.26±0.12(4) | 0.93 |
| O | 8.69 | 8.87±0.07(6) | 0.23 | 9.01±0.08(5) | 0.42 | 8.73±0.09(6) | 0.09 | 8.84±0.11(6) | 0.65 |
| Na | 6.24 | 6.51±0.07(6) | 0.32 | 6.30±0.10(3) | 0.16 | 6.53±0.06(4) | 0.34 | 5.95±0.03(4) | 0.21 |
| Mg (Mg I) | 7.60 | 7.67±0.08(5) | 0.12 | 7.60±0.07(5) | +0.10 | 7.41±0.06(4) | −0.14 | 7.15±0.07(3) | 0.05 |
| Mg (MgH) | $\cdots$ | 7.70 | 0.15 | 7.66 | +0.16 | 7.41 | −0.14 | 7.15 | 0.05 |
| Al | 6.45 | 6.64±0.08(4) | 0.24 | 6.41±0.06(4) | +0.06 | 6.38±0.10(4) | −0.02 | 6.19±0.06(2) | 0.24 |
| Si | 7.51 | 7.56±0.10(5) | 0.1 | 7.60±0.08(5) | 0.19 | 7.61±0.05(5) | 0.15 | 7.45±0.08(2) | 0.44 |
| Ca | 6.34 | 6.44±0.08(10) | 0.15 | 6.36±0.10(9) | +0.12 | 6.20±0.10(8) | −0.09 | 5.95±0.14(8) | 0.11 |
| Sc I | 3.15 | $\cdots$ | $\cdots$ | $\cdots$ | $\cdots$ | $\cdots$ | $\cdots$ | $\cdots$ | $\cdots$ |
| Sc II | 3.15 | 3.17±0.07(7) | 0.07 | 3.32±0.06(6) | +0.27 | 3.16±0.16(6) | 0.06 | 2.91±0.13(5) | 0.26 |
| Ti I | 4.95 | 5.02±0.08(13) | 0.12 | 5.04±0.07(11) | +0.19 | 4.91±0.11(9) | 0.01 | 4.77±0.08(5) | 0.32 |
| Ti II | 4.95 | 5.03±0.08(4) | 0.13 | 5.05±0.07(3) | 0.20 | 4.98±0.14(4) | 0.08 | 4.72±0.12(4) | 0.27 |
| V | 3.93 | 4.03±0.05(6) | 0.15 | 3.82±0.08(6) | -0.01 | 3.85±0.06(6) | −0.03 | 3.62±0.05(6) | 0.19 |
| Cr | 5.64 | 5.72±0.08(11) | 0.13 | 5.75±0.06(10) | 0.21 | 5.61±0.10(10) | 0.02 | 5.20±0.08(10) | 0.06 |
| Mn | 5.43 | 5.86±0.07(5) | 0.48 | 5.87±0.06(4) | +0.54 | 5.33±0.16(4) | −0.05 | 4.94±0.10(3) | 0.01 |
| Fe I | 7.50 | 7.45±0.09(49) | −0.05 | 7.40±0.10(39) | −0.10 | 7.45±0.07(22) | −0.05 | 7.00±0.10(22) | −0.50 |
| Fe II | $\cdots$ | 7.46±0.05(12) | −0.04 | 7.40±0.07(10) | −0.10 | 7.44±0.02(2) | −0.06 | 7.00±0.05(2) | −0.50 |
| Co | 4.99 | 4.97±0.07(6) | 0.03 | 4.96±0.06(6) | 0.07 | 4.96±0.12(6) | 0.02 | 4.73±0.12(6) | 0.24 |
| Ni | 6.22 | 6.29±0.07(6) | 0.12 | 6.16±0.08(5) | 0.04 | 6.09±0.13(5) | −0.08 | 5.83±0.14(5) | 0.11 |
| Zn I | 4.56 | 4.66±0.07(2) | 0.15 | 3.99(1) | −0.47 | 4.37±0.07(2) | −0.14 | 4.58±0.07(2) | 0.52 |
| La II | 1.10 | 1.73(1) | 0.68 | 1.70(1) | 0.7 | 1.54(1) | 0.49 | 1.09(1) | 0.49 |
| Eu II | 0.52 | $\cdots$ | $\cdots$ | $\cdots$ | $\cdots$ | 0.63(1) | −0.06 | 0.3(1) | 0.28 |

[a]This work

**HR 6766**

HR 6766 or HD 165634 is a well-studied weak CH(G)-band star (wGb-star). It is an RC star based on its position on the HR diagram. Our determination of stellar parameters (Table 2) and elemental abundance (Table 3) agree well with those reported by Palacios et al. (2012) and Adamczak & Lambert (2013). The weaker G-band results from a lower C-abundance. The derived $\log \epsilon(\text{Li})$ of 1.18 dex. The C I lines are very weak in the spectrum, and the derived $\log \epsilon(\text{C})$ is 7.45 dex. Since, the CN-bands are very weak the N-abundance from N I lines ($\log \epsilon(\text{N}) = 8.67 \pm 0.09$ dex) is adopted, and also yielding only the lower limit to the $^{12}\text{C}/^{13}\text{C} > 10$. This is an extreme case of mixing observed in giants with high depletion in C and enhancement in N. Despite the efficient mixing, Li is not destroyed and shows an abundance similar to the Sun ($\log \epsilon(\text{Li}) = 1.05$ dex). HR 6766 show no He-enhancement (refer to Figure 1, Table 10).

**KIC 9821622**

KIC 9821622 is a K-type giant, with a He-inert core, and a confirmed RGB-bump star by the Kepler's asteroseismic data. Our derived stellar parameters (Table 2) and elemental abundances (Table 3) agree well with those reported by Singh et al. (2020). The N abundance derived from both N I lines and the CN-band are same within the uncertainties and is enhanced as expected for giants. The He/H ratio derived from Mg-features (Figure 1) and CH-features (Figure 3) is 0.1 and He is not enhanced (Table 10).

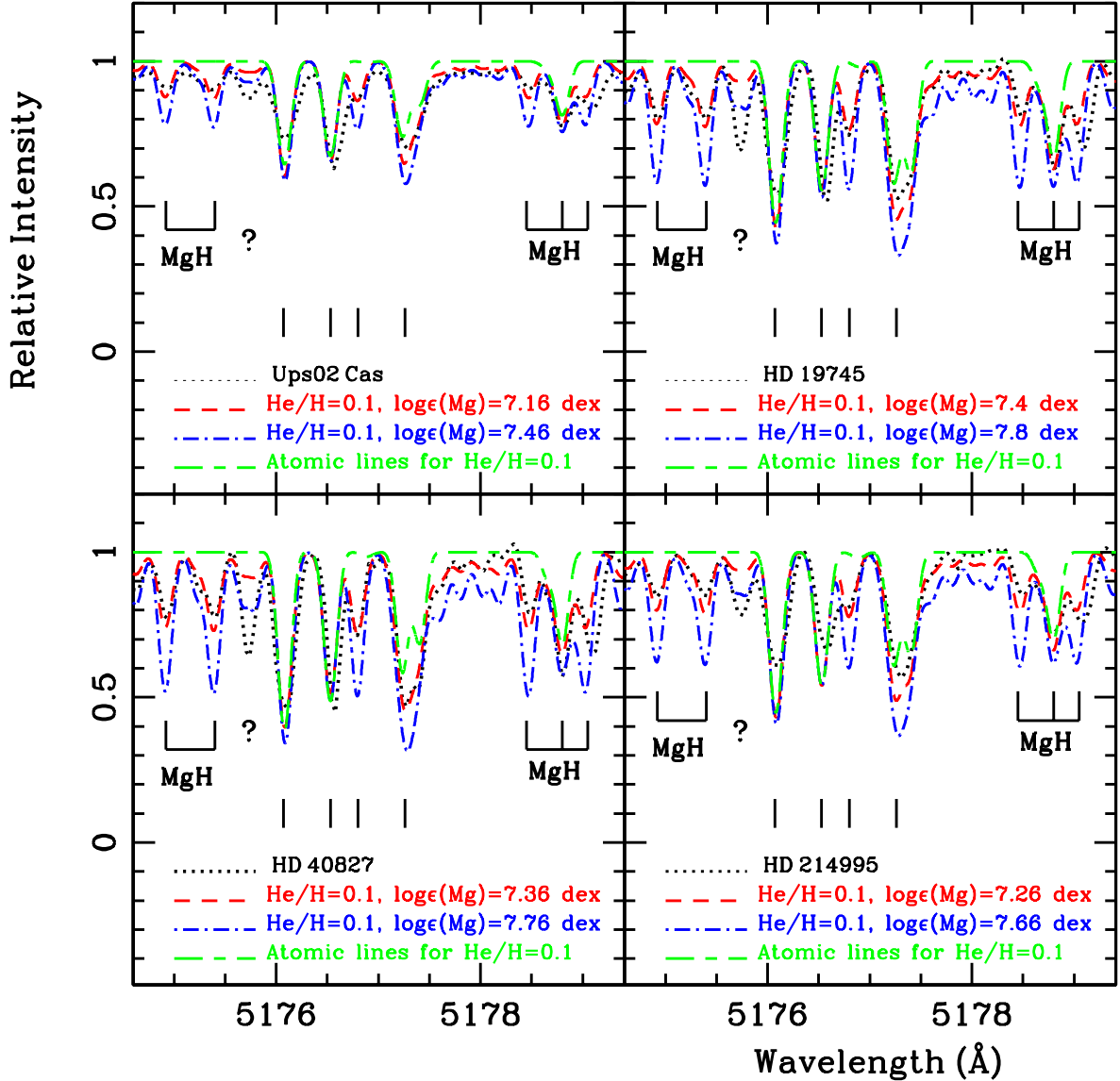

Fig. 5.— Figure show the observed and the synthetic spectra of MgH band. The dotted black line is the observed spectrum of the program star. The best fit synthesized spectrum is shown with dashed red line for the stars' derived He/H ratio (=0.1) and the corresponding Mg-abundance (from Mg I and MgH). The synthesis is also shown with dash dotted blue line for slightly higher Mg-abundance that is not providing the fit. The synthesis for pure atomic lines is shown with the long and short dashed green line. The MgH lines are marked. The vertical lines are the atomic lines, Co I, Ni I, V I and Fe I from blue to redward in the spectrum and the question mark is an unknown line.

Table 4.   Elemental abundances for program stars in the order of their decreasing $T_{\rm eff}$. The numbers given in parentheses with the abundances are the number of spectral lines of that species used for analysis.

| Elements | Sun | Ups02 Cas | | KIC 2305930 | | | | HD 19745 | |
|---|---|---|---|---|---|---|---|---|---|
| | $\log \epsilon(X)_\odot$ | $\log \epsilon(X)_{\rm He/H=0.1}$ | [X/Fe] | $\log \epsilon(X)_{\rm He/H=0.1}$ | [X/Fe] | $\log \epsilon(X)_{\rm He/H=0.4}$ | [X/Fe] | $\log \epsilon(X)_{\rm He/H=0.1}$ | [X/Fe] |
| H | 12.00 | 12.00 | $\cdots$ | 12.00 | $\cdots$ | 11.735 | $\cdots$ | 12.00 | $\cdots$ |
| He | 10.93 | 11.00 | $\cdots$ | 11.00 | $\cdots$ | 11.337 | $\cdots$ | 11.00 | $\cdots$ |
| Li | 1.05 | $< -0.3$ | $< -0.94$ | 4.03±0.08(2) | 3.36 | 3.79(2) | 3.49 | 4.55(1) | 3.5 |
| C | 8.43 | 8.11±0.08(4) | 0.08 | 8.11±0.04(3) | 0.06 | 7.32±0.1(3) | $-0.36$ | 8.5±0.15(3) | 0.07 |
| CH | 8.46 | 8.35±0.15(3) | 0.29 | 8.11±0.04(3) | 0.03 | 8.05±0.05(3) | 0.34 | 8.47±0.06(3) | 0.01 |
| $^{12}C/^{13}C$ [a] | 89 | 15±1 | $\cdots$ | 6±1 | $\cdots$ | 6±1 | $\cdots$ | 15[b] | $\cdots$ |
| N(CN) | 7.83 | 7.7±0.05 | 0.27 | 7.8±0.10 | +0.35 | 7.55 | 0.46 | 7.4±0.06 | $-0.43$ |
| O | 8.69 | 8.72±0.1(6) | 0.43 | 9.5±0.10(5) | 1.19 | 8.81±0.1(5) | +0.86 | 8.6±0.14(6) | $-0.09$ |
| Na | 6.24 | 5.80±0.07(4) | $-0.04$ | 5.83±0.10(3) | $-0.03$ | 5.63±0.11(3) | 0.14 | 6.21±0.10(4) | $-0.03$ |
| Mg (Mg I) | 7.60 | 7.26±0.06(5) | 0.06 | 7.70±0.08(4) | $-0.48$ | 7.39±0.08(4) | +0.54 | 7.40±0.09(4) | $-0.2$ |
| Mg (MgH) | $\cdots$ | 7.16 | $-0.04$ | 7.20 | $-0.02$ | 7.39 | +0.54 | 7.40 | $-0.2$ |
| Al | 6.45 | 6.09±0.06(4) | 0.04 | 6.29±0.10(4) | +0.22 | 6.00±0.13(4) | +0.3 | 6.38±0.08(4) | $-0.07$ |
| Si | 7.51 | 7.23±0.05(5) | 0.12 | 7.72±0.08(5) | 0.59 | 7.32±0.07(5) | 0.56 | 7.65±0.11(5) | 0.14 |
| Ca | 6.34 | 5.85±0.06(9) | $-0.09$ | 5.97±0.09(10) | -0.01 | 5.68±0.14(10) | $-0.09$ | 6.16±0.12(8) | $-0.18$ |
| Sc I | 3.15 | $\cdots$ | $\cdots$ | $\cdots$ | $\cdots$ | $\cdots$ | $\cdots$ | $\cdots$ | $\cdots$ |
| Sc II | 3.15 | 2.94±0.10(7) | 0.19 | 3.03±0.11(5) | 0.26 | 2.75±0.13(5) | +0.35 | 3.05±0.15(6) | +0.1 |
| Ti I | 4.95 | 4.43±0.08(13) | $-0.12$ | 4.43±0.11(11) | $-0.14$ | 4.31±0.10(10) | +0.11 | 4.83±0.09(9) | $-0.12$ |
| Ti II | 4.95 | 4.41±0.09(4) | $-0.14$ | 4.43±0.05(4) | $-0.14$ | 4.17±0.11(4) | $-0.03$ | 4.77±0.05(4) | $-0.18$ |
| V | 3.93 | 3.38±0.03(6) | $-0.15$ | 3.54±0.10(6) | $-0.01$ | 3.39±0.12(6) | 0.21 | 3.97±0.11(6) | +0.04 |
| Cr | 5.64 | 5.01±0.09(11) | $-0.23$ | 5.65±0.10(5) | +0.39 | 5.43±0.12(5) | 0.00 | 5.60±0.14(10) | $-0.04$ |
| Mn | 5.43 | 4.86±0.07(4) | $-0.17$ | 5.29±0.11(4) | +0.24 | 5.13±0.10(4) | 0.45 | 5.38±0.12(4) | $-0.05$ |
| Fe I | 7.50 | 7.09±0.08(49) | $-0.41$ | 7.12±0.11(41) | $-0.38$ | 6.89±0.13(41) | $-0.61$ | 7.50±0.13(20) | 0.0 |
| Fe II | 7.50 | 7.10±0.07(12) | $-0.40$ | 7.12±0.09(10) | $-0.38$ | 6.64±0.09(10) | $-0.86$ | 7.50±0.06(2) | 0.0 |
| Co | 4.99 | 4.49±0.08(7) | $-0.1$ | 4.72±0.11(6) | 0.11 | 4.55±0.11(6) | 0.31 | 5.08±0.08(6) | 0.09 |
| Ni | 6.22 | 5.82±0.08(5) | 0.00 | 6.09±0.11(5) | +0.25 | 5.82±0.12(5) | 0.35 | 6.18±0.16(5) | $-0.04$ |
| Zn I | 4.56 | 4.49±0.03(2) | 0.33 | 4.69±0.03(2) | 0.51 | 4.26±0.03(2) | +0.45 | $\cdots$ | $\cdots$ |
| La II | 1.10 | 1.05(1) | 0.35 | 0.8(1) | 0.08 | 0.66(1) | 0.16 | 1.18(1) | 0.08 |

[a] This work

[b] de La Reza & Drake (1995); de La Reza & da Silva (1995a); da Silva et al. (1995)

**Ups02 Cas, HD 19745, HD 40827, and HD 214995**

Ups02 Cas is a K-giant having no Li with only the upper limit of $\log \epsilon(\text{Li}) = < -0.3$ dex (see Table 4). Our determination of stellar parameters and the metallicity are in good agreement with those reported by Charbonnel & Balachandran (2000) and references therein. For HD 19745 is at RGB-bump phase by its position on the HR diagram with $\log \epsilon(\text{Li}) = 4.55$ dex. Our derived stellar parameters (Table 2), and elemental abundances (Table 4) are in agreement with those reported by Reddy & Lambert (2005) and de La Reza & da Silva (1995b), within the uncertainties. HD 40827 is an RGB star with $\log \epsilon(\text{Li}) = 1.87$ dex. The derived stellar parameters and the elemental abundances (Table 5) agree well with those reported by Gratton & D'Antona (1989). HD 214995 is a giant star at its RGB-bump phase by its position on the HR diagram with $\log \epsilon(\text{Li}) = 3.27 \pm 0.03$ dex. The derived stellar parameters and the elemental abundance (Table 5) are in excellent agreement with those reported by Kumar et al. (2011) and Luck & Heiter (2007).

For all these stars derived He/H ratio is 0.1 (Figure 5) with no enhancement in He (Table 10).

**HDE 233517, HR 334 and HD 4813**

HDE 233517 is a super Li-rich giant at the RGB-bump phase with $\log \epsilon(\text{Li}) = 4.24$ dex. The derived stellar parameters and elemental abundances (Table 5) are in good agreement with those reported by Balachandran et al. (2000). HR 334 is at RGB-bump phase with $\log \epsilon(\text{Li}) = 1.37$ dex. The derived stellar parameters and elemental abundances (Table 6) agrees with Luck & Challener (1995) and Lambert et al. (1980) within the uncertainties. HD 4813 is a K-giant at its RGB bump phase having very a weak/no detectable Li line, with an upper limit of $\log \epsilon(\text{Li}) < -0.2$ dex. The derived stellar parameters, metallicity and Li-abundances (Table 6) agree well with those reported by Lambert et al. (1980). HR 334 and HD 4813 are at the RGB-bump phase but the Li-abundances observed are different. The derived He/H ratio is 0.1 for these program stars (Figure 6) with no He-enhancement (Table 10).

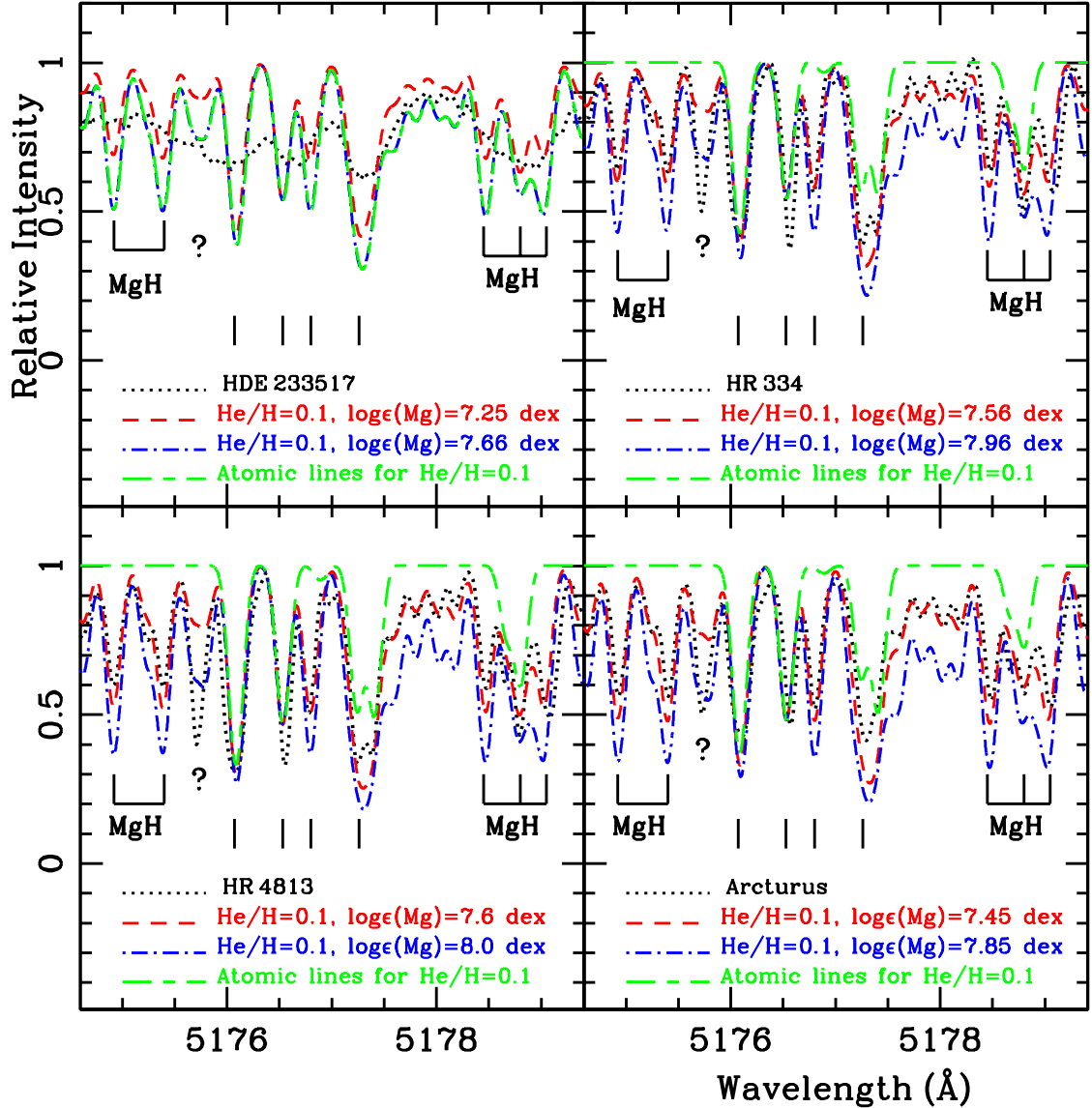

Fig. 6.— Figure show the observed and the synthetic spectra of MgH band. The dotted black line is the observed spectrum of the program star. The best fit synthesized spectrum is shown with dashed red line for the stars' derived He/H ratio (=0.1) and the corresponding Mg-abundance (from Mg I and MgH). The synthesis is also shown with dash dotted blue line for slightly higher Mg-abundance that is not providing the fit. The synthesis for pure atomic lines is shown with the long and short dashed green line. The MgH lines are marked. The vertical lines are the atomic lines, Co I, Ni I, V I and Fe I from blue to redward in the spectrum and the question mark is an unknown line.

Table 5. Elemental abundances for program stars in the order of their decreasing $T_{\text{eff}}$. The numbers given in parentheses with the abundances are the number of spectral lines of that species used for analysis.

| Elements | Sun $\log \epsilon(\text{X})_\odot$ | HD 40827 $\log \epsilon(\text{X})_{\text{He/H}=0.1}$ | [X/Fe] | HDE 233517 $\log \epsilon(\text{X})_{\text{He/H}=0.1}$ | [X/Fe] | HD 214995 $\log \epsilon(\text{X})_{\text{He/H}=0.1}$ | [X/Fe] |
|---|---|---|---|---|---|---|---|
| H | 12.00 | 12.00 | $\cdots$ | 12.00 | $\cdots$ | 12.00 | $\cdots$ |
| He | 10.93 | 11.00 | $\cdots$ | 11.00 | $\cdots$ | 11.00 | $\cdots$ |
| Li | 1.05 | 1.87 | 0.92 | 4.24±0.06(2) | 3.44 | 3.27±0.03(2) | 2.32 |
| C | 8.43 | 8.47±0.06(4) | 0.14 | 8.25(1) | 0.07 | 8.5±0.1(1) | 0.17 |
| CH | 8.46 | 8.60±0.15(3) | 0.24 | 8.3 | 0.09 | 8.5±0.05(3) | 0.17 |
| $^{12}$C/$^{13}$C$^a$ | 89 | 15±1 | $\cdots$ | >20±2 | $\cdots$ | 13 | $\cdots$ |
| N(CN) | 7.83 | 8.30±0.1 | 0.57 | 8.1 | 0.52 | 8.15±0.1 | 0.22 |
| O | 8.60 | 9.07±0.11(6) | 0.48 | 9.34±0.10(5) | 0.9 | 8.95±0.1(3) | 0.36 |
| Na | 6.24 | 6.37±0.10(4) | 0.23 | 5.96±.04(2) | -0.03 | 6.22±0.18(4) | 0.08 |
| Mg (Mg I) | 7.60 | 7.52±0.08(5) | 0.02 | 7.24±0.06(2) | -0.11 | 7.26±0.11(3) | -0.24 |
| Mg (MgH) | 7.60 | 7.36 | -0.14 | 7.25 | -0.10 | 7.26 | -0.24 |
| Al | 6.45 | 6.48±0.06(4) | 0.13 | 6.64±0.11 | 0.44 | 6.26±0.05(2) | -0.09 |
| Si | 7.51 | 7.58±0.12(5) | 0.17 | $\cdots$ | $\cdots$ | 7.70±0.10(5) | 0.29 |
| Ca | 6.34 | 6.15±0.08(10) | -0.09 | 5.97±0.12(6) | -0.12 | 6.11±0.13(6) | -0.13 |
| Sc I | 3.15 | $\cdots$ | $\cdots$ | | | 3.26±0.13(4) | -0.11 |
| Sc II | 3.15 | 3.08±0.09(7) | 0.03 | 3.61±0.11(7) | 0.71 | 3.23±0.11(4) | 0.11 |
| Ti I | 4.95 | 4.80±0.12(13) | -0.05 | 4.96±0.08(10) | 0.26 | 4.70±0.07(4) | -0.15 |
| Ti II | 4.95 | 4.70±0.10(4) | -0.15 | 5.00±0.08(3) | 0.30 | 4.70±0.16(4) | -0.24 |
| V | 3.93 | 3.85±0.09(6) | 0.02 | 3.98±0.09(5) | 0.30 | 3.98±0.14(4) | 0.15 |
| Cr | 5.64 | 5.34±0.09(11) | -0.2 | 5.24±0.11(7) | -0.15 | 5.47±0.11(4) | -0.07 |
| Mn | 5.43 | 5.48±0.13(5) | +0.15 | 5.98±0.04(2) | -0.8 | 5.39±0.14(3) | 0.06 |
| Fe I | 7.50 | 7.40±0.11(48) | -0.10 | 7.25±0.10(36) | -0.25 | 7.4±0.11(19) | -0.10 |
| Fe II | $\cdots$ | 7.41±0.09(12) | -0.09 | 7.25±0.11(9) | -0.25 | 7.40±0.14(7) | -0.10 |
| Co | 4.99 | 5.10±0.13(7) | 0.21 | 4.63±0.06(4) | -0.11 | 5.03±0.13(6) | 0.14 |
| Ni | 6.22 | 6.27±0.07(6) | 0.15 | 6.23±0.10(4) | 0.26 | 6.23±0.09(3) | 0.11 |
| Zn I | 4.56 | 4.54±0.12(2) | 0.08 | 3.79(1) | -0.52 | $\cdots$ | $\cdots$ |
| La II | 1.10 | 1.71(1) | 0.71 | 1.36(1) | 0.51 | 1.74(1) | 0.74(1) |

**α Boo, HD 205349, HD 39853**

α Boo or Arcturus is a standard RGB star with no Li spectral line in its observed spectrum yielding only the upper limit of $\log \epsilon(\text{Li}) < -1.0$ dex, as expected for giants. Our derived stellar parameters (Table 2) and the elemental abundances (Table 7) are in good agreement with those reported by Lambert & Ries (1981) and Ramírez & Allende Prieto (2011), within the uncertainties.

HD 205349 is a probable supergiant star with the derived $\log \epsilon(\text{Li}) = 1.92$ dex. The derived stellar parameters and elemental abundances (Table 7) are in good agreement with those reported by Gonçalves et al. (2020) and Brown et al. (1989), within the uncertainties. The observed La abundance of [La/Fe]=0.92 dex is as expected for a supergiant star. And, HD 39853 is a super Li-rich early-AGB star with Li-abundance of $\log \epsilon(\text{Li}) = 3.92$ dex. The derived stellar parameters and elemental abundances (Table 9) are in good agreement with those reported by Gratton & D'Antona (1989).

The derived He/H ratio for these program stars is 0.1 (Figures 6 and 7) with no He-enhancement (Table 10).

## 6.2. He-enhanced giants in the order of decreasing $T_{\text{eff}}$

**KIC 2305930**

KIC 2305930 is an RC star from the Kepler field. This is a super Li-rich giant with $\log \epsilon(\text{Li}) = 4.03$ dex (see Table 4). From the studies of Jorissen et al. (2020), KIC 2305930 is a binary star, with v sin$i$=13 kms$^{-1}$. The derived stellar parameters and the abundances are in excellent agreement with those reported by (Kumar et al. 2018).

The Mg-abundance derived from the MgH band is about 0.5 dex less than that derived from the Mg I lines, $\log \epsilon(\text{Mg})$=7.70±0.08 dex (see Figure 8). Further, the MgH band synthesized by adopting a model with a He/H ratio of 0.4 and $[\text{Fe/H}]_{He}$[7]$= -0.6$, provided the best fit to the observed MgH band for the $\log \epsilon(\text{Mg I}) = 7.40$ dex, which is same as that derived from the Mg I lines. For the pair of He/H ratio and [Fe/H] the elemental abundances were derived (see Table 4). The derived Mg abundance from Mg-features and C abundance from C-features for different He/H ratios are given in Table 10.

**HD 107484**

HD 107484 is at the tip of the RGB based on its position on the HR diagram, with a Li-rich abundance of $\log \epsilon(\text{Li}) = 2.10$ dex. Our derived stellar parameter and elemental abundances

---

[7][Fe/H]$_{He}$ denotes the metallicity of the He-enhanced giants determined for He-rich models (He/H>0.1), that derived for the star.

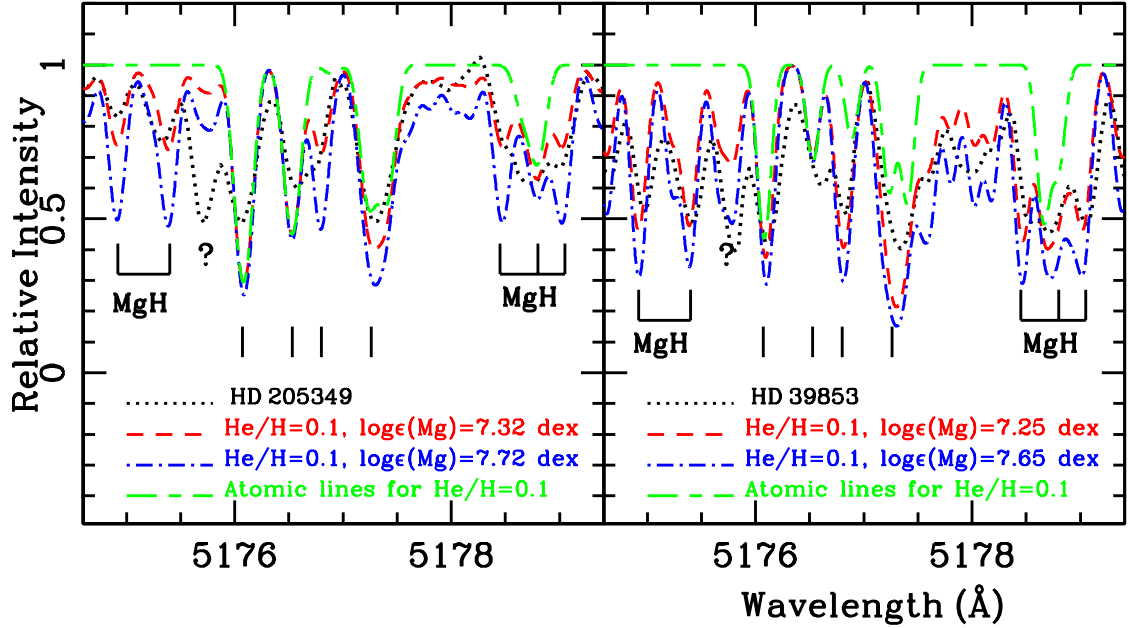

Fig. 7.— Figure show the observed and the synthetic spectra of MgH band. The dotted black line is the observed spectrum of the program star. The best fit synthesized spectrum is shown with dashed red line for the stars' derived He/H ratio (=0.1) and the corresponding Mg-abundance (from Mg I and MgH). The synthesis is also shown with dash dotted blue line for slightly higher Mg-abundance that is not providing the fit. The synthesis for pure atomic lines is shown with the long and short dashed green line. The MgH lines are marked. The vertical lines are the atomic lines, Co I, Ni I, V I and Fe I from blue to redward in the spectrum and the question mark is an unknown line.

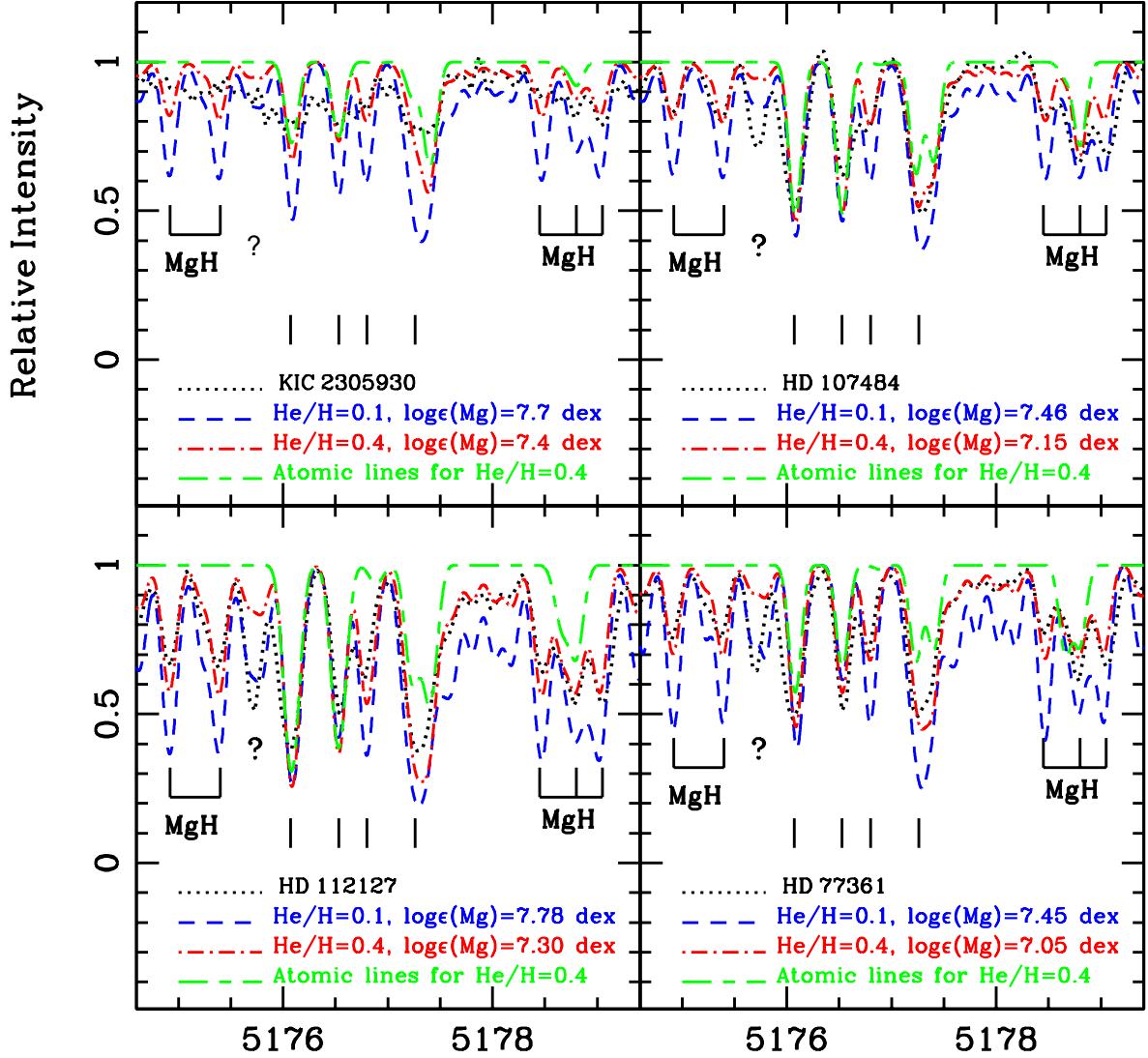

Fig. 8.— Figure show the observed and the synthetic spectra of MgH band. The dotted black line is the observed spectrum of the program star. The dash dotted blue line is the synthetic spectrum for the normal He/H ratio of 0.1 and the corresponding Mg-abundance (from Mg I). The dashed red line shows the synthetic spectrum for the derived, enhanced He/H ratio, and the corresponding Mg-abundance (from Mg I and MgH) that is providing the best fit. The synthesis for pure atomic lines is shown with the long and short dashed green line. The MgH lines are marked. The vertical lines are the atomic lines, Co I, Ni I, V I and Fe I from blue to redward in the spectrum and the question mark is an unknown line.

(Table 6) agree well with those reported by Kumar et al. (2011) within the uncertainties. The Mg-abundance derived from the MgH band is 0.36 dex less than that derived from the Mg I lines, $\log \epsilon(\text{Mg})$=7.46 dex (see Figure 8). Further, the MgH band synthesized for the adopted model with a He/H ratio of 0.4 with $[\text{Fe/H}]_{He} = -0.5$, provided the best fit to the observed MgH band for the $\log \epsilon(\text{Mg})$=7.08 dex, which is same as that derived from the Mg I lines (Table 10). For this pair of He/H ratio and [Fe/H], all the elemental abundances are derived (Table 6).

Table 6. Elemental abundances for program stars in the order of their decreasing $T_{\text{eff}}$. The numbers given in parentheses with the abundances are the number of spectral lines of that species used for analysis.

| Elements | Sun $\log\epsilon(\text{X})_\odot$ | HD 107484 | | | | HR 334 | | HR 4813 | |
|---|---|---|---|---|---|---|---|---|---|
| | | $\log\epsilon(\text{X})_{\text{He/H}=0.1}$ | [X/Fe] | $\log\epsilon(\text{X})_{\text{He/H}=0.4}$ | [X/Fe] | $\log\epsilon(\text{X})_{\text{He/H}=0.1}$ | [X/Fe] | $\log\epsilon(\text{X})_{\text{He/H}=0.1}$ | [X/Fe] |
| H | 12.00 | 12.00 | $\cdots$ | 11.735 | $\cdots$ | 12.00 | $\cdots$ | 12.00 | $\cdots$ |
| He | 10.93 | 11.00 | $\cdots$ | 11.372 | $\cdots$ | 11.00 | $\cdots$ | 11.00 | $\cdots$ |
| Li | 1.05 | 2.10±0.07(2) | 1.20 | 1.80±0.12(2) | 1.25 | 1.37(1) | 0.28 | < −0.2 | < −0.9 |
| C | 8.43 | 8.44±0.09(5) | 0.16 | 7.54±0.13(5) | -0.39 | 8.65±0.1(2) | 0.18 | 8.7±0.06(2) | 0.19 |
| C(CH) | 8.46 | 8.50±0.1(3) | 0.19 | 8.50±0.1(3) | 0.54 | 8.65±0.04(3) | 0.2 | 8.75±0.05(3) | 0.21 |
| $^{12}$C/$^{13}$C | 89 | 20±1 | $\cdots$ | 20±1 | $\cdots$ | 20[b] | $\cdots$ | 20[b] | $\cdots$ |
| N(CN) | 7.83 | 8.1 | 0.42 | 7.7 | 0.37 | 7.55 | −0.32 | 7.55 | −0.33 |
| O | 8.69 | 9.04±0.09(6) | 0.50 | 8.35±0.2(6) | 0.16 | 8.85±0.12(3) | +0.12 | 8.84±0.05(3) | 0.1 |
| Na | 6.24 | 6.48±0.07(4) | 0.39 | 6.19±0.08(4) | 0.45 | 6.46±0.13(4) | 0.18 | 6.63±0.12(4) | 0.34 |
| Mg (Mg I) | 7.60 | 7.46±0.02(5) | 0.01 | 7.08±0.13(5) | −0.02 | 7.56±0.05(3) | −0.08 | 7.61±0.08(3) | −0.04 |
| Mg (Mg H) | $\cdots$ | 7.10 | −0.35 | 7.05 | −0.05 | 7.56 | −0.08 | 7.61 | −0.04 |
| Al | 6.45 | 6.50±0.07(4) | 0.20 | 6.14±0.04(4) | 0.19 | 6.47±0.05(2) | −0.02 | 6.58±0.06(2) | 0.08 |
| Si | 7.51 | 7.49±0.11(5) | 0.13 | 7.06±0.11(5) | 0.05 | 7.73±0.11(5) | 0.18 | 7.87±0.09(5) | 0.31 |
| Ca | 6.34 | 6.18±0.09(10) | −0.01 | 5.78±0.15(10) | −0.06 | 6.26±0.12(8) | −0.12 | 6.20±0.08(8) | −0.19 |
| Sc I | 3.15 | $\cdots$ | $\cdots$ | $\cdots$ | $\cdots$ | $\cdots$ | $\cdots$ | $\cdots$ | $\cdots$ |
| Sc II | 3.15 | 3.06±0.12(7) | 0.06 | 2.74±0.14(7) | 0.09 | 3.11±0.15(6) | −0.08 | 3.19±0.10(6) | −0.01 |
| Ti I | 4.95 | 4.80±0.06(12) | 0.00 | 4.57±0.09(12) | +0.12 | 4.90±0.11(8) | −0.09 | 4.97±0.08(8) | −0.03 |
| Ti II | 4.95 | 4.72±0.07(4) | −0.08 | 4.33±0.07(4) | −0.12 | 4.85±0.07(3) | −0.14 | 4.92±0.05(3) | −0.08 |
| V | 3.93 | 4.04±0.08(6) | 0.26 | 3.71±0.10(6) | 0.28 | 4.08±0.06(6) | 0.11 | 4.15±0.10(6) | 0.17 |
| Cr | 5.64 | 5.49±0.09(10) | 0.00 | 5.23±0.12(10) | +0.09 | 5.68±0.11(10) | 0.00 | 5.71±0.11(10) | 0.02 |
| Mn | 5.43 | 5.39±0.01(4) | 0.11 | 5.15±0.11(4) | 0.22 | 5.57±0.12(4) | 0.1 | 5.58±0.05(4) | 0.10 |
| Fe I | 7.50 | 7.35±0.09(48) | −0.15 | 7.0±0.15(48) | −0.50 | 7.54±0.13(21) | 0.04 | 7.55±0.11(21) | 0.05 |
| Fe II | $\cdots$ | 7.37±0.07(11) | −0.13 | 6.80±0.12(11) | −0.70 | 7.55±0.02(2) | 0.05 | 7.54±0.01(2) | 0.04 |
| Co | 4.99 | 4.90±0.13(7) | 0.06 | 4.65±0.12(7) | 0.16 | 5.23±0.14(6) | 0.2 | 5.34±0.13(6) | 0.30 |
| Ni | 6.22 | 6.41±0.10(4) | 0.34 | 6.04±0.2(4) | +0.32 | 6.28±0.15(5) | 0.02 | 6.39±0.12(5) | 0.12 |
| Zn | 4.56 | 4.47±0.10(2) | 0.06 | 3.93±0.18(2) | −0.13 | $\cdots$ | $\cdots$ | $\cdots$ | $\cdots$ |
| La II | 1.10 | 1.54(1) | 0.59 | 1.27(1) | 0.67 | 0.98(1) | −0.16 | 0.87(1) | −0.28 |

[b]Luck & Challener (1995)

**HD 112127**

HD 112127 is the first Li-rich giant discovered by Wallerstein & Sneden (1982) by the presence of a strong Li-spectral line in its observed spectrum. The derived Li abundance is $\log \epsilon$ (Li)=3.4 dex. The evolutionary state was determined by asteroseismology using the light curves extracted from TESS FFI [8] and it is found to be in RC-phase. The derived stellar parameters and the elemental abundances (Table 7) are in good agreement with those reported by Wallerstein & Sneden (1982); Brown et al. (1989); Luck & Challener (1995) within the uncertainties.

The Mg-abundance derived from the MgH band is 0.48 dex less than that derived from the Mg I lines, that is $\log \epsilon$(Mg)=7.78 dex (see Figure 8). Further, the MgH band synthesized by adopting a model with a He/H ratio of 0.4 and $[Fe/H]_{He} = -0.4$, provided the best fit to the observed MgH band for $\log \epsilon$(Mg)=7.30 dex, which is same as that derived from the Mg I lines (Table 10). For this pair of He/H ratio and [Fe/H] all the elemental abundances are derived (Table 7).

---

[8]https://mast.stsci.edu/portal/Mashup/Clients/Mast/Portal.html

Table 7. Elemental abundances for program stars in the order of their decreasing $T_{\rm eff}$. The numbers given in parentheses with the abundances are the number of spectral lines of that species used for analysis.

| Elements | Sun $\log \epsilon(X)_\odot$ | α Boo $\log \epsilon(X)_{\rm He/H=0.1}$ | [X/Fe] | HD 205349 $\log \epsilon(X)_{\rm He/H=0.1}$ | [X/Fe] | HD 112127 $\log \epsilon(X)_{\rm He/H=0.1}$ | [X/Fe] | $\log \epsilon(X)_{\rm He/H=0.4}$ | [X/Fe] |
|---|---|---|---|---|---|---|---|---|---|
| H | 12.00 | 12.00 | ⋯ | 12.00 | ⋯ | 12.00 | ⋯ | 11.735 | ⋯ |
| He | 10.93 | 11.00 | ⋯ | 11.00 | ⋯ | 11.00 | ⋯ | 11.337 | ⋯ |
| Li | 1.05 | < −1.0 | <1.43 | 1.92(1) | 1.12 | 3.40±0.10(2) | 2.28 | 3.10±0.1(2) | 2.45 |
| C | 8.43 | 8.15±0.07(3) | 0.34 | 8.6±0.05 | 0.42 | 8.8±0.14(2) | 0.30 | 7.65±0.07(2) | −0.38 |
| C(CH) | 8.46 | 8.18±0.02(3) | +0.34 | 8.5±0.1 | 0.32 | 8.60±0.1(3) | 0.07 | 8.60±0.1(3) | 0.54 |
| $^{12}$C/$^{13}$C | 89±1 | 8±1 | ⋯ | ⋯ | ⋯ | 22[a] | ⋯ | ⋯ | ⋯ |
| N | 7.83 | 7.45±0.05(CN) | +0.24 | 7.80±0.05 | +0.22 | 7.85±0.1 | −0.05 | 7.45 | 0.02 |
| O | 8.69 | 8.75±0.04(5) | 0.68 | 8.85±0.08 | 0.41 | 8.95±0.02(3) | 0.19 | 8.63±0.06(3) | 0.34 |
| Na | 6.24 | 5.77±0.08(4) | +0.15 | 6.57±0.15(3) | 0.58 | 6.78±0.18(3) | 0.47 | 6.43±0.07(3) | 0.59 |
| Mg (Mg I) | 7.60 | 7.48±0.10(5) | +0.50 | 7.22±0.04(3) | −0.13 | 7.78±0.05(2) | 0.11 | 7.32±0.05(2) | 0.12 |
| Mg (MgH) | ⋯ | 7.45 | +0.47 | 7.32 | −0.03 | 7.30 | −0.37 | 7.30 | 0.1 |
| Al | 6.45 | 6.23±0.10(4) | 0.40 | ⋯ | ⋯ | 6.14±0.0(2) | −0.38 | 5.84±0.04(2) | −0.21 |
| Si | 7.51 | 7.20±0.08(5) | 0.31 | 7.50±0.14(5) | 0.24 | 7.79±0.19(4) | 0.21 | 7.27±0.16(4) | 0.13 |
| Ca | 6.34 | 5.93±0.07(10) | +0.21 | 5.69±0.16(6) | −0.40 | 6.19±0.15(6) | −0.22 | 5.83±0.18(6) | −0.11 |
| Sc I | 3.15 | 2.74±0.10(4) | +0.21 | 3.23±0.11(4) | +0.33 | 3.62±0.07(2) | 0.40 | 3.39±0.03(2) | 0.64 |
| Sc II | 3.15 | 2.75±0.07(7) | +0.22 | 3.27±0.15(5) | +0.37 | 3.12±0.11(5) | −0.1 | 2.75±0.10(5) | 0.00 |
| Ti I | 4.95 | 4.64±0.09(13) | +0.31 | 4.69±0.12(5) | −0.01 | 4.94±0.13(6) | −0.08 | 4.73±0.13(6) | 0.18 |
| Ti II | 4.95 | 4.65±0.12(4) | +0.32 | 4.62±0.12(4) | −0.08 | 4.88±0.18(2) | −0.14 | 4.40±0.14(2) | −0.15 |
| V | 3.93 | 3.57±0.09(6) | +0.26 | 3.86±0.10(4) | +0.18 | 4.28±0.13(3) | 0.28 | 4.03±0.13(3) | 0.50 |
| Cr | 5.64 | 5.01±0.12(11) | −0.01 | 5.36±0.17(5) | −0.03 | 5.50±0.15(6) | −0.21 | 5.21±0.15(6) | −0.03 |
| Mn | 5.43 | 4.79±0.12(5) | −0.02 | 5.07±0.13(3) | −0.11 | 5.29±0.15(3) | −0.21 | 5.03±0.19(3) | +0.13 |
| Fe I | 7.50 | 6.88±0.05(50) | −0.62 | 7.25±0.10(15) | −0.25 | 7.57±0.16(16) | 0.07 | 7.10±0.13(16) | −0.40 |
| Fe II | ⋯ | 6.88±0.07(12) | −0.62 | 7.25±0.09(5) | −0.25 | 7.57±0.19(6) | 0.07 | 6.95±0.11(6) | −0.55 |
| Co | 4.99 | 4.74±0.13(7) | 0.37 | 4.82±0.10(6) | 0.08 | 5.41±0.16(4) | 0.35 | 5.02±0.18(4) | 0.29 |
| Ni | 6.22 | 5.76±0.11(6) | 0.16 | 6.10±0.10(3) | +0.13 | 6.54±0.15(2) | 0.25 | 6.06±0.12(2) | 0.24 |
| Zn I | 4.56 | 4.20±0.0(2) | 0.26 | ⋯ | ⋯ | ⋯ | ⋯ | ⋯ | ⋯ |
| La | 1.10 | 1.74(1) | 0.74 | 1.77(1) | 0.92 | 1.75(1) | 0.58 | 1.41(1) | 0.71 |

[a]Berdyugina & Savanov (1994); de La Reza & Drake (1995); Brown et al. (1989); Wallerstein & Sneden (1982)

**HD 77361**

HD 77361 is an RGB-bump star based on its position on the HR diagram with super Li-rich abundance of $\log\epsilon$ (Li) = 3.9 dex. The determination of stellar parameters and elemental abundances (Table 8) is carried out by Hema & Pandey (2020), and these are in excellent agreement with that reported by Lyubimkov et al. (2015).

The Mg-abundance derived from the MgH band is 0.45 dex less than that derived from the Mg I lines, $\log\epsilon$(Mg)=7.45 dex (see Figure 8). Further, the MgH band synthesized by adopting a model with a He/H ratio of 0.4 and [Fe/H]$_{He}$ = $-0.3$, provided the best fit to the observed MgH band for the $\log\epsilon$(Mg)=7.05 dex, which is the same as that derived from the Mg I lines (Table 10). For the adopted pair of He/H ratio and the [Fe/H], all the elemental abundances are derived (Table 8).

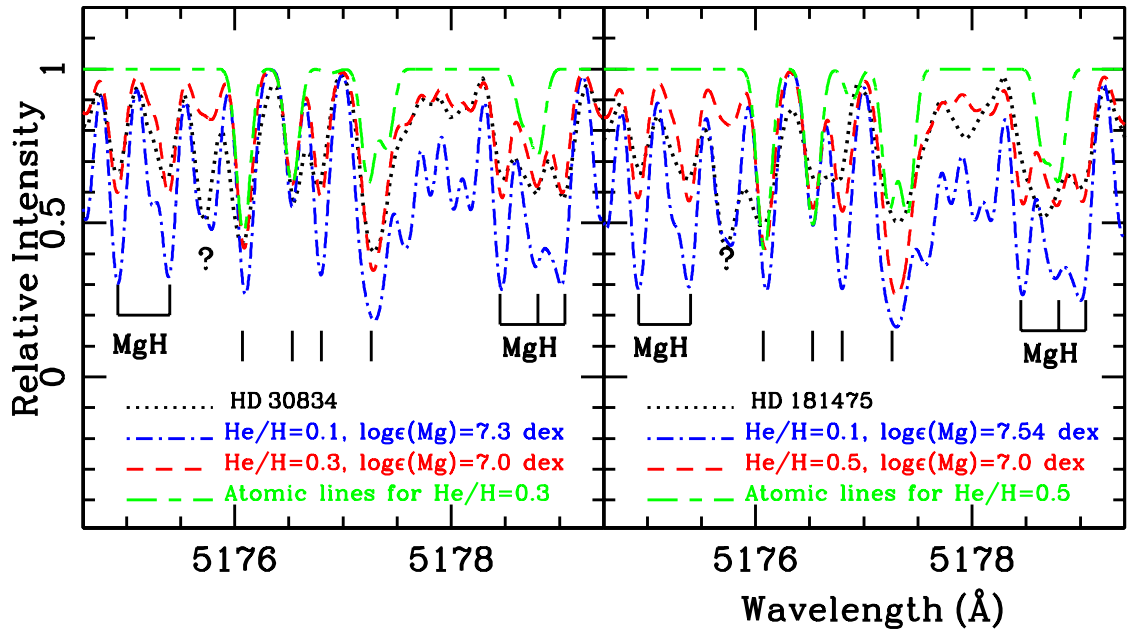

Fig. 9.— Figure show the observed and the synthetic spectra of MgH band. The dotted black line is the observed spectrum of the program star. The dash dotted blue line is the synthetic spectrum for the normal He/H ratio of 0.1 and the corresponding Mg-abundance (from Mg I). The dashed red line shows the synthetic spectrum for the derived, enhanced He/H ratio, and the corresponding Mg-abundance (from Mg I and MgH) that is providing the best fit. The synthesis for pure atomic lines is shown with the long and short dashed green line. The MgH lines are marked. The vertical lines are the atomic lines, Co I, Ni I, V I and Fe I from blue to redward in the spectrum and the question mark is an unknown.

Table 8.   Elemental abundances for program stars in the order of their decreasing $T_{\rm eff}$. The numbers given in parentheses with the abundances are the number of spectral lines of that species used for analysis.

| Elements | Sun $\log \epsilon(\mathrm{X})_{\odot}$ | HD 77361 $\log \epsilon(\mathrm{X})_{\mathrm{He/H=0.1}}$ | [X/Fe] | $\log \epsilon(\mathrm{X})_{\mathrm{He/H=0.4}}$ | [X/Fe] | HD 30834 $\log \epsilon(\mathrm{X})_{\mathrm{He/H=0.1}}$ | [X/Fe] | $\log \epsilon(\mathrm{X})_{\mathrm{He/H=0.3}}$ | [X/Fe] |
|---|---|---|---|---|---|---|---|---|---|
| H | 12.00 | 12.00 | ⋯ | 11.735 | ⋯ | 12.00 | ⋯ | 11.735 | ⋯ |
| He | 10.93 | 11.00 | ⋯ | 11.337 | ⋯ | 11.00 | ⋯ | 11.337 | ⋯ |
| Li | 1.05 | 3.90(1) | 2.9 | 3.67(1) | 3.02 | 2.3±0.1(2) | 1.43 | 2.1±0.1(2) | 1.50 |
| C | 8.43 | 8.7±0.14(3) | 0.32 | 7.8±0.02(3) | −0.23 | 8.18(1) | −0.07 | 7.55(1) | −0.43 |
| C(CH) | 8.46 | 8.35±0.1(3) | −0.06 | 8.30±0.1(3) | 0.24 | 8.25±0.1(3) | −0.03 | 8.30±0.1(3) | 0.29 |
| $^{12}$C/$^{13}$C | 89±1 | 8±1 | ⋯ | 8±1 | ⋯ | 13$^{\mathrm{a}}$ | ⋯ | ⋯ | ⋯ |
| N(CN) | 7.83 | 7.40 | −0.38 | 7.25 | −0.18 | 7.05±0.1 | −0.6 | 6.85±0.1 | −0.53 |
| O | 8.69 | 8.75±0.02(2) | 0.06 | 8.45±0.03(2) | 0.16 | 8.5±0.07(2) | 0.04 | 8.41±0.07(2) | 0.18 |
| Na | 6.24 | 6.18±0.05(3) | −0.01 | 5.89±0.05(3) | 0.05 | 6.29±0.17(4) | 0.23 | 6.0±0.13(4) | 0.21 |
| Mg (Mg I) | 7.60 | 7.45±0.13(4) | −0.10 | 7.05±0.09(4) | −0.15 | 7.28±0.1(3) | −0.14 | 7.0±0.06(3) | −0.15 |
| Mg (MgH) | ⋯ | 7.00 | −0.55 | 7.05 | −0.15 | 7.05 | −0.37 | 7.0 | −0.15 |
| Al | 6.45 | 6.40±0.08(4) | 0.00 | 6.05±0.08(4) | 0.00 | 6.18±0.14(2) | −0.09 | 5.84±0.05(2) | −0.16 |
| Si | 7.51 | 7.80±0.12(5) | 0.34 | 7.30±0.11(5) | 0.19 | 7.47±0.1(2) | 0.14 | 7.20±0.06(2) | 0.14 |
| Ca | 6.34 | 5.93±0.18(8) | −0.36 | 5.58±0.17(8) | −0.36 | 6.15±0.14(7) | −0.01 | 5.72±0.13(7) | −0.17 |
| Sc I | 3.15 | 2.81±0.13(3) | −0.29 | 2.69±0.12(3) | −0.06 | 3.10±0.07(4) | 0.13 | 2.73±0.07(4) | 0.03 |
| Sc II | 3.15 | 3.00±0.12(5) | −0.10 | 2.69±0.07(5) | −0.06 | 2.90±0.1(4) | −0.07 | 2.80±0.13(4) | 0.1 |
| Ti I | 4.95 | 4.48±0.05(6) | −0.42 | 4.31±0.06(6) | −0.24 | 4.78±0.07(4) | 0.01 | 4.40±0.1(4) | −0.1 |
| Ti II | 4.95 | 4.62±0.08(4) | −0.28 | 4.24±0.10(4) | −0.31 | 4.73±0.14(3) | −0.04 | 4.46±0.14(3) | −0.04 |
| V | 3.93 | 3.56±0.10(4) | −0.32 | 3.34±0.10(4) | −0.19 | 3.68±0.11(4) | −0.07 | 3.25±0.1(4) | −0.23 |
| Cr | 5.64 | 5.32±0.14(7) | −0.27 | 5.03±0.12(7) | −0.21 | 5.17±0.16(5) | −0.29 | 4.80±0.19(5) | −0.39 |
| Mn | 5.43 | 5.30±0.20(3) | −0.08 | 5.06±0.19(3) | −0.03 | 5.18±0.16(3) | −0.07 | 4.94±0.15(3) | −0.04 |
| Fe I | 7.50 | 7.45±0.12(18) | −0.05 | 7.10±0.12(18) | −0.40 | 7.32±0.16(15) | −0.18 | 7.05±0.16(15) | −0.45 |
| Fe II | ⋯ | 7.50±0.05(4) | 0.00 | 7.00±0.05(4) | −0.50 | 7.32±0.14(2) | −0.18 | 7.00±0.14(2) | −0.5 |
| Co | 4.99 | 5.13±0.17(6) | 0.19 | 4.78±0.17(6) | 0.19 | 4.89±0.14(6) | 0.08 | 4.64±0.09(6) | 0.1 |
| Ni | 6.22 | 6.19±0.18(3) | 0.02 | 5.81±0.13(3) | −0.01 | 5.86±0.09(2) | −0.18 | 5.65±0.07(2) | −0.12 |
| La | 1.10 | 2.10(1) | 1.05 | 1.76(1) | 1.06 | 2.02(1) | 1.1 | 1.73(1) | 1.08 |

$^{\mathrm{a}}$Berdyugina & Savanov (1994); de La Reza & Drake (1995)

**HD 30834**

HD 30834 is an early-AGB star from its position on the HR diagram with Li-rich abundance of $\log \epsilon(\text{Li}) = 2.3$ dex. The stellar parameters (Table 2) and the elemental abundances (Table 8) derived are in excellent agreement with those reported by Cayrel de Strobel et al. (1997) and Brown et al. (1989).

The Mg-abundance derived from the MgH band is about 0.3 dex less than that derived from the Mg I lines, $\log \epsilon(\text{Mg}) = 7.28$ dex (see Figure 9). Further, the MgH band synthesized by adopting a model for a pair of He/H ratio=0.3 and $[\text{Fe/H}]_{He} = -0.45$, provided the best fit to the observed MgH band for the $\log \epsilon(\text{Mg}) = 7.00$ dex, which is same as that derived from the Mg I lines (Table 10). For this pair of He/H and [Fe/H], the elemental abundances are derived (Table 8).

**HD 181475**

HD 181475 is a super giant that has no or very little Li in its atmosphere, yielding only the upper limit of $\log \epsilon(\text{Li}) < -0.8$ dex, which is in agreement with that reported by Drake et al. (2002). For the adopted model atmosphere (Table 2) the elemental abundances are derived (Table 9).

The Mg-abundance derived from the MgH band is about 0.54 dex less than that derived from the Mg I lines, $\log \epsilon(\text{Mg}) = 7.54$ dex (see Figure 9). Further, the MgH band synthesized by adopting a model with a He/H ratio of 0.5 and $[\text{Fe/H}]_{He} = -0.3$, provided the best fit to the observed MgH band for the $\log \epsilon(\text{Mg}) = 7.00$ dex, which is same as that derived from the Mg I lines (Table 10). For the adopted pair of He/H and [Fe/H], the elemental abundances are derived (Table 9).

Table 9.  Elemental abundances for program stars in the order of their decreasing $T_{\rm eff}$. The numbers given in parentheses with the abundances are the number of spectral lines of that species used for analysis.

| | Sun | HD 181475 | | | | HD 39853 | |
|---|---|---|---|---|---|---|---|
| Elements | $\log \epsilon(\mathrm{X})_\odot$ | $\log \epsilon(\mathrm{X})_{\mathrm{He/H}=0.1}$ | [X/Fe] | $\log \epsilon(\mathrm{X})_{\mathrm{He/H}=0.5}$ | [X/Fe] | $\log \epsilon(\mathrm{X})_{\mathrm{He/H}=0.1}$ | [X/Fe] |
| H | 12.00 | 12.00 | $\cdots$ | 11.673 | $\cdots$ | 12.00 | $\cdots$ |
| He | 10.93 | 11.00 | $\cdots$ | 11.372 | $\cdots$ | 11.00 | $\cdots$ |
| Li | 1.05 | $< -0.8(1)$ | $-1.96$ | $-1.1(1)$ | $-1.83$ | 3.90±0.03(2) | 3.05 |
| C | 8.43 | 8.7(1) | 0.16 | 7.70(1) | -0.41 | 8.50(1) | 0.27 |
| C(CH) | 8.46 | 8.35±0.1 | $-0.22$ | 8.33±0.1 | 0.19 | 8.5±0.1 | 0.24 |
| $^{12}\mathrm{C}/^{13}\mathrm{C}$ | 89±1 | $\cdots$ | $\cdots$ | $\cdots$ | $\cdots$ | 6[a] | $\cdots$ |
| N(CN) | 7.83 | 7.40±0.1 | $-0.54$ | 7.20±0.1 | $-0.31$ | 7.5 | $-0.13$ |
| O | 8.69 | 9.1±0.06(2) | 0.30 | 8.76±0.07(2) | 0.39 | 8.53±0.05(2) | 0.04 |
| Na | 6.24 | 6.63±0.17(4) | 0.28 | 6.30±0.25(4) | 0.38 | 6.20±0.18(4) | 0.16 |
| Mg (Mg I) | 7.60 | 7.54±0.01(3) | $-0.17$ | 7.03±0.03(3) | $-0.25$ | 7.25±0.04(3) | $-0.15$ |
| Mg (MgH) | $\cdots$ | 7.00 | $-0.71$ | 7.00 | $-0.28$ | 7.28 | $-0.12$ |
| Al | 6.45 | 6.08±0.11(2) | $-0.48$ | 5.75±0.04(2) | $-0.38$ | 6.27±0.07(2) | 0.02 |
| Si | 7.51 | 7.85±0.14(4) | 0.23 | 7.16±0.14(4) | $-0.03$ | 7.39±0.1(5) | 0.08 |
| Ca | 6.34 | 5.55±0.19(5) | $-0.9$ | 5.20±0.17(5) | $-0.82$ | 6.13±0.12(6) | $-0.01$ |
| Sc I | 3.15 | $\cdots$ | $\cdots$ | $\cdots$ | $\cdots$ | 3.02±0.04(2) | 0.07 |
| Sc II | 3.15 | 3.41±0.16(6) | 0.15 | 3.01±0.17(6) | 0.18 | 3.00±0.06(3) | 0.05 |
| Ti I | 4.95 | 4.89±0.12(4) | $-0.17$ | 4.6±0.12(4) | $-0.03$ | 5.31±0.1(4) | 0.56 |
| Ti II | 4.95 | 4.89±0.12(4) | $-0.17$ | 4.5±0.10(4) | $-0.13$ | 5.17±0.03(3) | 0.42 |
| V | 3.93 | 4.35±0.17(6) | 0.31 | 4.18±0.2(6) | 0.57 | 4.52±0.09(4) | 0.80 |
| Cr | 5.64 | 5.42±0.14(6) | $-0.33$ | 5.22±0.11(6) | $-0.1$ | 5.55±0.13(4) | 0.11 |
| Mn | 5.43 | 5.63±0.12(3) | 0.09 | 5.41±0.15(3) | 0.30 | 5.11±0.13(3) | $-0.12$ |
| Fe I | 7.50 | 7.61±0.16(13) | 0.11 | 7.18±0.19(13) | $-0.32$ | 7.30±0.15(16) | $-0.20$ |
| Fe II | $\cdots$ | $\cdots$ | $\cdots$ | $\cdots$ | $\cdots$ | 7.28(1) | $-0.20$ |
| Co | 4.99 | 5.03±0.15(4) | $-0.07$ | 4.68±0.14(4) | 0.01 | 5.05±0.12(4) | 0.26 |
| Ni | 6.22 | 6.47±0.16(3) | 0.14 | 5.90±0.19(3) | 0.00 | 5.59±0.03(3) | $-0.43$ |
| La | 1.10 | 2.63(1) | 1.42 | 2.28(1) | 1.5 | 1.16(1) | 0.26 |

[a]da Silva et al. (1995)

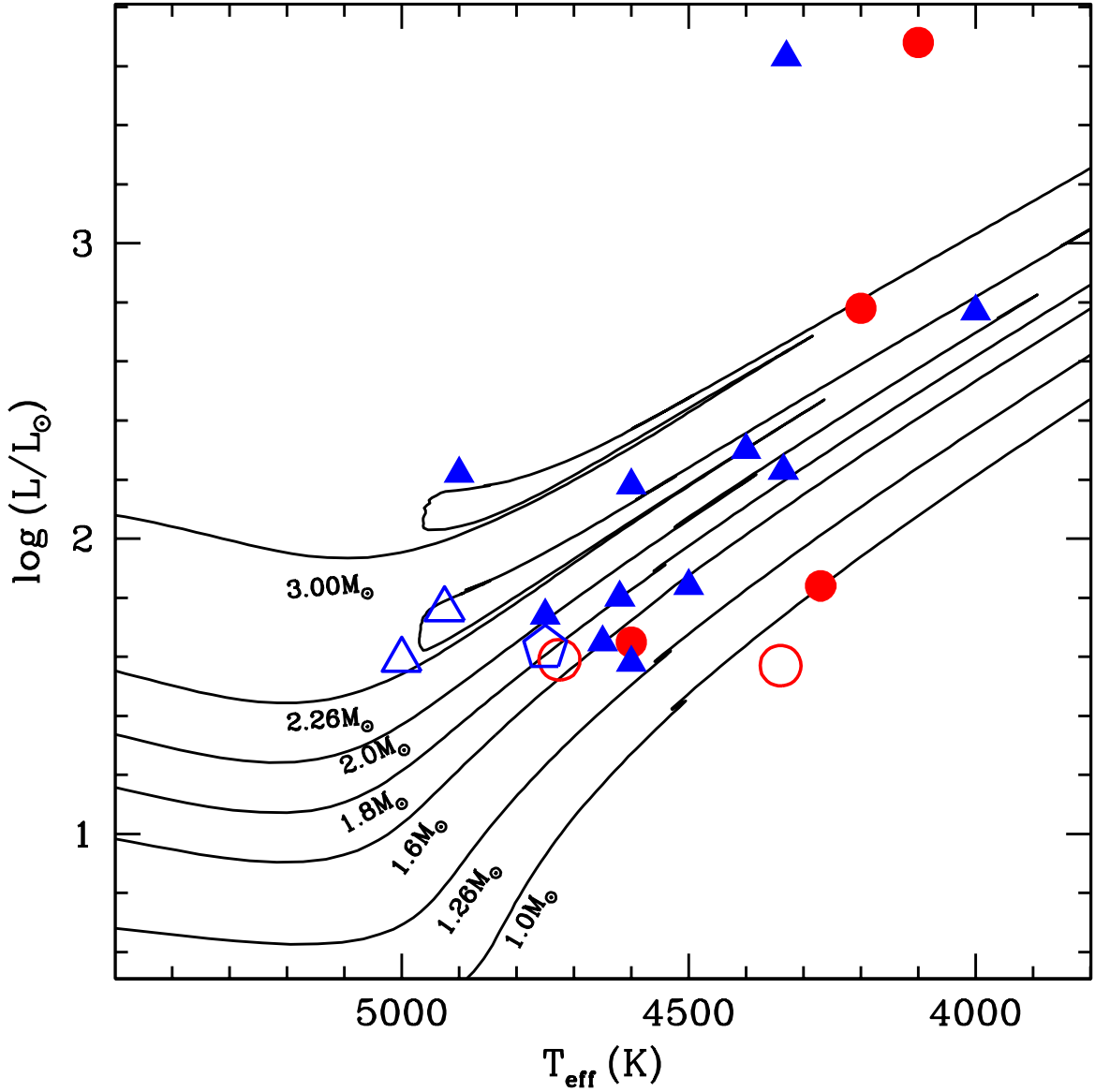

Fig. 10.— Figure shows HR-diagram for the program stars – the Luminosity (L⊙) versus the $T_{\rm eff}$. The blue solid triangles are the He-normal giants, red dots are the He-enriched giants. The open triangles are the He-normal RC-stars and red circles are the He-enriched RC-stars from Kepler data. The open pentagon is an RGB-bump star: KIC 9821622 from Kepler data.

Table 10. The comparison of log ε(Mg) derived from the Mg I and MgH band and, the log ε(C) derived from the C I and CH-band (G-band) from the three regions.

| Stars | He/H | log ε(Mg I) | log ε(MgH) | log ε(CI) | log ε(CH[a]) | log ε(CH[b]) | log ε(CH[c]) |
|---|---|---|---|---|---|---|---|
| KIC 5184199 | 0.1 | 7.67 | 7.70 | 8.27 | 8.40 | 8.40 | 8.40 |
| | 0.2 | 7.52 | 7.72 | 8.09 | 8.40 | 8.40 | 8.40 |
| | 0.3 | 7.45 | 7.70 | 7.92 | 8.35 | 8.35 | 8.35 |
| | 0.4 | 7.36 | 7.72 | 7.68 | 8.40 | 8.40 | 8.40 |
| | 0.5 | 7.27 | 7.77 | 7.40 | 8.35 | 8.35 | 8.35 |
| KIC 12645107 | 0.1 | 7.60 | 7.66 | 7.88 | 7.95 | 7.88 | 7.90 |
| | 0.2 | 7.42 | 7.70 | 7.55 | 7.85 | 7.85 | 7.80 |
| | 0.3 | 7.32 | 7.70 | 7.23 | 7.80 | 7.80 | 7.85 |
| | 0.4 | 7.26 | 7.65 | 7.10 | 7.90 | 7.90 | 7.85 |
| | 0.5 | 7.21 | 7.70 | 6.90 | 7.85 | 7.85 | 7.95 |
| HR 6766 | 0.1 | 7.41 | 7.41 | 7.45 | >7.00 | >7.00 | >7.0 |
| | 0.2 | 7.26 | 7.45 | 7.24 | >7.70 | >7.70 | >7.70 |
| | 0.3 | 7.15 | 7.40 | 7.04 | >7.70 | >7.70 | >7.70 |
| | 0.4 | 7.07 | 7.45 | 6.90 | >7.65 | >7.65 | >7.65 |
| | 0.5 | 7.01 | 7.45 | 6.68 | >7.60 | >7.60 | >7.60 |
| KIC 9821622 | 0.1 | 7.15 | 7.15 | 8.15 | 8.20 | 8.20 | 8.17 |
| | 0.2 | 7.04 | 7.20 | 7.94 | 8.20 | 8.15 | 8.20 |
| | 0.3 | 6.95 | 7.15 | 7.68 | 8.15 | 8.20 | 8.20 |
| | 0.4 | 6.89 | 7.15 | 7.56 | 8.20 | 8.15 | 8.15 |
| | 0.5 | 6.84 | 7.20 | 7.48 | 8.20 | 8.20 | 8.25 |
| Ups02 Cas | 0.1 | 7.26 | 7.16 | 8.10 | 8.30 | 8.35 | 8.35 |
| | 0.2 | 7.20 | 7.10 | 7.85 | 8.40 | 8.30 | 8.30 |
| | 0.3 | 7.00 | 7.20 | 7.72 | 8.35 | 8.35 | 8.40 |
| | 0.4 | 6.92 | 7.20 | 7.55 | 8.30 | 8.30 | 8.35 |
| | 0.5 | 6.90 | 7.15 | 7.37 | 8.35 | 8.30 | 8.35 |
| KIC 2305930 | 0.1 | 7.70 | 7.20 | 8.10 | 8.05 | 8.10 | 8.15 |
| | 0.2 | 7.55 | 7.25 | 7.79 | 8.10 | 8.10 | 8.15 |
| | 0.3 | 7.45 | 7.30 | 7.55 | 8.10 | 8.10 | 8.20 |
| | 0.4 | 7.39 | 7.39 | 7.32 | 8.00 | 8.05 | 8.10 |

| Stars | He/H | log $\epsilon$(Mg I) | log $\epsilon$(MgH) | log $\epsilon$(CI) | log $\epsilon$(CH[a]) | log $\epsilon$(CH[b]) | log $\epsilon$(CH[c]) |
|---|---|---|---|---|---|---|---|
| | 0.5 | 7.34 | 7.40 | 7.23 | 8.10 | 8.10 | 8.15 |
| HD 19745 | 0.1 | 7.40 | 7.40 | 8.50 | 8.47 | 8.47 | 8.47 |
| | 0.2 | 7.23 | 7.40 | 8.27 | 8.50 | 8.50 | 8.50 |
| | 0.3 | 7.10 | 7.45 | 7.90 | 8.45 | 8.45 | 8.45 |
| | 0.4 | 7.03 | 7.40 | 7.80 | 8.45 | 8.45 | 8.45 |
| | 0.5 | 6.97 | 7.40 | 7.70 | 8.40 | 8.45 | 8.40 |
| HD 40827 | 0.1 | 7.52 | 7.36 | 8.47 | 8.60 | 8.60 | 8.65 |
| | 0.2 | 7.45 | 7.35 | 8.30 | 8.55 | 8.50 | 8.60 |
| | 0.3 | 7.28 | 7.40 | 8.10 | 8.65 | 8.55 | 8.55 |
| | 0.4 | 7.20 | 7.40 | 7.95 | 8.60 | 8.55 | 8.50 |
| | 0.5 | 7.10 | 7.40 | 7.80 | 8.60 | 8.55 | 8.55 |
| HD 214995 | 0.1 | 7.26 | 7.26 | 8.50 | 8.50 | 8.50 | 8.50 |
| | 0.2 | 7.10 | 7.30 | 8.30 | 8.45 | 8.45 | 8.40 |
| | 0.3 | 7.00 | 7.30 | 8.16 | 8.50 | 8.45 | 8.45 |
| | 0.4 | 6.92 | 7.28 | 8.05 | 8.45 | 8.50 | 8.50 |
| | 0.5 | 6.86 | 7.25 | 7.80 | 8.45 | 8.45 | 8.50 |
| HDE 233517 | 0.1 | 7.25 | 7.25 | 8.25 | 8.30 | 8.30 | 8.25 |
| | 0.2 | 7.15 | 7.20 | 8.10 | 8.25 | 8.20 | 8.20 |
| | 0.3 | 7.08 | 7.25 | 7.95 | 8.28 | 8.20 | 8.15 |
| | 0.4 | 6.92 | 7.25 | 7.83 | 8.20 | 8.15 | 8.23 |
| | 0.5 | 6.85 | 7.25 | 7.70 | 8.30 | 8.20 | 8.25 |
| HD 107484 | 0.1 | 7.46 | 7.10 | 8.44 | 8.55 | 8.50 | 8.44 |
| | 0.2 | 7.30 | 7.10 | 8.11 | 8.52 | 8.45 | 8.54 |
| | 0.3 | 7.16 | 7.00 | 7.86 | 8.50 | 8.45 | 8.54 |
| | 0.4 | 7.08 | 7.05 | 7.54 | 8.50 | 8.50 | 8.54 |
| | 0.5 | 6.85 | 6.95 | 7.35 | 8.52 | 8.42 | 8.54 |
| HR 334 | 0.1 | 7.56 | 7.56 | 8.65 | 8.65 | 8.65 | 8.65 |
| | 0.2 | 7.37 | 7.56 | 8.34 | 8.65 | 8.65 | 8.65 |
| | 0.3 | 7.24 | 7.56 | 8.03 | 8.55 | 8.55 | 8.55 |

| Stars | He/H | log $\epsilon$(Mg I) | log $\epsilon$(MgH) | log $\epsilon$(CI) | log $\epsilon$(CH[a]) | log $\epsilon$(CH[b]) | log $\epsilon$(CH[c]) |
|-------|------|------|------|------|------|------|------|
| | 0.4 | 7.18 | 7.45 | 7.82 | 8.56 | 8.56 | 8.56 |
| | 0.5 | 7.11 | 7.45 | 7.62 | 8.55 | 8.55 | 8.55 |
| HR 4813 | 0.1 | 7.61 | 7.61 | 8.70 | 8.75 | 8.75 | 8.75 |
| | 0.2 | 7.41 | 7.60 | 8.38 | 8.75 | 8.75 | 8.75 |
| | 0.3 | 7.28 | 7.60 | 8.05 | 8.75 | 8.75 | 8.75 |
| | 0.4 | 7.21 | 7.60 | 8.04 | 8.65 | 8.65 | 8.65 |
| | 0.5 | 7.14 | 7.60 | 7.93 | 8.65 | 8.65 | 8.65 |
| HD 112127 | 0.1 | 7.78 | 7.30 | 8.80 | 8.55 | 8.60 | 8.55 |
| | 0.2 | 7.60 | 7.30 | 8.55 | 8.60 | 8.55 | 8.55 |
| | 0.3 | 7.45 | 7.25 | 7.95 | 8.55 | 8.60 | 8.55 |
| | 0.4 | 7.32 | 7.30 | 7.65 | 8.55 | 8.60 | 8.55 |
| | 0.5 | 7.15 | 7.35 | 7.40 | 8.50 | 8.60 | 8.55 |
| $\alpha$ Boo | 0.1 | 7.48 | 7.45 | 8.15 | 8.18 | 8.18 | 8.18 |
| | 0.2 | 7.30 | 7.43 | 8.00 | 8.16 | 8.16 | 8.16 |
| | 0.3 | 7.15 | 7.42 | 7.87 | 8.13 | 8.13 | 8.13 |
| | 0.4 | 7.00 | 7.42 | 7.72 | 8.13 | 8.13 | 8.13 |
| | 0.5 | 6.80 | 7.40 | 7.50 | 8.12 | 8.12 | 8.12 |
| HD 205349 | 0.1 | 7.22 | 7.32 | 8.60 | 8.50 | 8.50 | 8.50 |
| | 0.2 | 7.03 | 7.25 | 8.20 | 8.45 | 8.45 | 8.45 |
| | 0.3 | 6.94 | 7.25 | 7.90 | 8.35 | 8.40 | 8.35 |
| | 0.4 | 6.86 | 7.30 | 7.70 | 8.35 | 8.35 | 8.35 |
| | 0.5 | 6.80 | 7.25 | 7.60 | 8.40 | 8.35 | 8.35 |
| HD 77361 | 0.1 | 7.45 | 7.00 | 8.70 | 8.40 | 8.35 | 8.40 |
| | 0.2 | 7.30 | 7.10 | 8.40 | 8.35 | 8.35 | 8.40 |
| | 0.3 | 7.15 | 7.05 | 8.10 | 8.35 | 8.30 | 8.35 |
| | 0.4 | 7.05 | 7.05 | 7.80 | 8.30 | 8.30 | 8.35 |
| | 0.5 | 6.92 | 7.05 | 7.60 | 8.25 | 8.35 | 8.30 |
| HD 30834 | 0.1 | 7.28 | 7.05 | 8.18 | 8.25 | 8.25 | 8.25 |
| | 0.2 | 7.16 | 7.00 | 7.80 | 8.20 | 8.20 | 8.20 |

| Stars | He/H | log $\epsilon$(Mg I) | log $\epsilon$(MgH) | log $\epsilon$(CI) | log $\epsilon$(CH[a]) | log $\epsilon$(CH[b]) | log $\epsilon$(CH[c]) |
|---|---|---|---|---|---|---|---|
| | 0.3 | 7.00 | 7.00 | 7.55 | 8.25 | 8.30 | 8.25 |
| | 0.4 | 6.87 | 6.96 | 7.30 | 8.25 | 8.25 | 8.25 |
| | 0.5 | 6.71 | 6.96 | 7.10 | 8.25 | 8.25 | 8.20 |
| HD 181475 | 0.1 | 7.54 | 7.00 | 8.70 | 8.35 | 8.35 | 8.40 |
| | 0.2 | 7.40 | 7.10 | 8.35 | 8.30 | 8.30 | 8.30 |
| | 0.3 | 7.27 | 7.10 | 8.15 | 8.30 | 8.30 | 8.30 |
| | 0.4 | 7.16 | 7.06 | 7.95 | 8.25 | 8.30 | 8.25 |
| | 0.5 | 7.03 | 7.00 | 7.70 | 8.33 | 8.33 | 8.33 |
| HD 39853 | 0.1 | 7.25 | 7.28 | 8.50 | 8.50 | 8.50 | 8.50 |
| | 0.2 | 6.97 | 7.25 | 8.20 | 8.45 | 8.45 | 8.45 |
| | 0.3 | 6.87 | 7.20 | 7.95 | 8.50 | 8.50 | 8.50 |
| | 0.4 | 6.77 | 7.25 | 7.75 | 8.40 | 8.40 | 8.40 |
| | 0.5 | 6.69 | 7.20 | 7.52 | 8.45 | 8.45 | 8.45 |

[a]The synthesis of CH-band in the wavelength window 4300-4308Å.

[b]The synthesis of CH-band in the wavelength window 4308-4315Å.

[c]The synthesis of CH-band in the wavelength window 4322-4327Å.

## 7. Results and Discussion

We aim to measure the He-enhancement in cool giants with a range in their surface Li-abundance from no/minimal to super Li (about ten times more than ISM Li). These giants are mostly in the red-giant branch that is from the RGB-bump, giant branch, tip of the giant branch and the RC phase. Two of our sample stars are early-AGB and two are supergiant stars that are analyzed for comparison (Figure 10 of this work and see Figure 1 of Iben (1967) for an illustrative HR diagram of stellar evolution).

In the MS phase, the nuclear reactions that convert hydrogen to helium in the core are: the proton-proton-chain (pp-chain) reaction and the CN-cycle. Many of our program stars (Table 2) are of about a solar mass ($M_\odot$) with a few exceptions, in which both pp-chain and CN-cycle operations are expected. The He produced in these reactions and the trace elements such as C, N, etc., that experience abundance changes including the ensuing isotopic ratios are brought to the surface during this dredge-up in the RGB phase. The surface Li is diluted during the mixing phase of sub giant and first-ascent red giant evolution. As a result, Li is not expected on the surface of the giants as the temperature involved during the mixing/dredge-up phase is sufficient to destroy Li (Iben 1967).

From many recent studies, it is observed that about ∼1.0% of the known giants are Li-rich (Deepak & Reddy 2022; Martell et al. 2021; Cai et al. 2023; Sitnova et al. 2023; Gao et al. 2019; Smiljanic et al. 2018). There could be one or more processes responsible for the Li-enrichment in giants such as preserving the pristine material with which the star is formed, the fresh production, or the contribution of external sources such as Li-rich planet engulfment. We expect that, in the Li-rich giants, if the Li is brought to the surface during the first dredge up on the giant branch, it also should be accompanied by He including other key elements. Hence, we expect to find such stars from our sample that are Li- as well as He-enriched. From our analyses, six stars were found to be having a large discrepancy in the derived Mg-abundances from the Mg I and MgH lines. And, that upon syntheses proved to be having higher He/H ratios than the standard value of 0.1 (See Section 5) along with super Li-rich abundances (see Figure 11). The Mg-features are the best to determine He/H ratios in giants. However, for the main-sequence stars like the Sun, the C-features provide more reliable values of He/H ratio (Moharana et al. 2024).

Along with the four Kepler field giants in our sample; KIC 9821622, KIC 12645107, KIC 2305930, and KIC 5184199, an effort was also made to derive the evolutionary states for other program stars using TESS FFI light curves. Except for HD 112127, which is an RC star (see Figure 12), the evolutionary state couldn't be derived for any other due to the incomplete data sets or with low signal-to-noise ratio. To distinguish between the RGB and the RC giants de la Reza (2025) has used the log $g$ values derived using asteroseismic parameters. By adopting the criteria that, for the RC-giants: $2.2 \leq \log g \leq 2.7$, and for RGB stars: $2.8 \leq \log g \leq 3.5$. Since the spectroscopically derived log $g$ values are in good agreement with those determined from seismic parameters, based on the spectroscopic log $g$ the evolutionary states are predicted for our

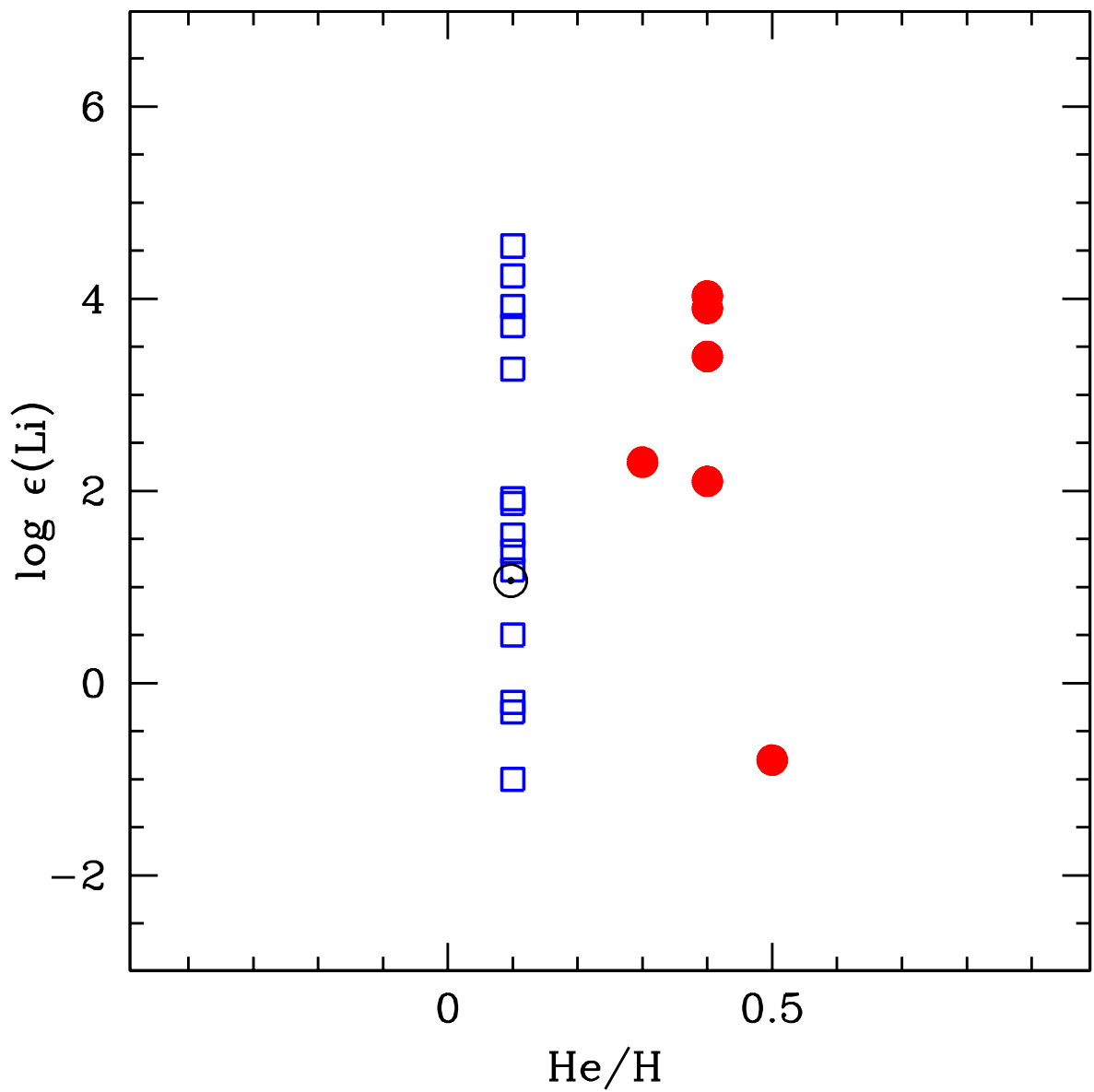

Fig. 11.— Figure shows the Li-abundance, log ε(Li) for the program stars versus the (He/H). The open blue squares are the stars with normal He/H ratios (He/H=0.1) and the filled red circles are the He-enhanced ones (He/H>0.1).

He-enhanced giants.

## 7.1. Results

### 7.1.1. He-rich stars

The results are discussed in the light of Li, He and CNO abundances, for program stars among other available elemental abundances. KIC 2305930 is an RC star with the derived He/H and [Fe/H]$_{He}$ ratios of 0.4 and $-0.6$, respectively, the star is not only He-enriched but also a super Li-rich with $\log \epsilon$ (Li)=4.03 dex. The abundance ratios are [C/Fe]=0.06, and [N/Fe]=+0.35 and the $^{12}$C/$^{13}$C ratio=6. The observed O-abundance ratio is about [O/Fe]=+1.19 dex. For giants the observed [O/Fe] is about +0.4 ((Sneden 1991) and references therein), and [O/Fe] of up to +1.0 has been observed for dwarfs (Abia & Rebolo 1989) of solar and near solar metallicity, and possibly increasing with decreasing metallicity. Here, the measured O-abundance plausibly represents the pristine material's O-abundance from which the star is formed. Table 11 shows the observed abundance of CNO, and the $\Delta$ here denotes the difference between the observed and the predicted initial O-abundance for the star's metallicity. If only the CN-cycle is operating, then the predicted N-abundance for the initial C and N-abundances is given by $(C+N)_0$, and if the CN- and ON-cycle is operating, the N-abundance predicted for the initial composition of C, N, and O are given by $(C+N+O)_0$. The observed N-abundance for KIC 2305930 is about 7.8 dex, and this is close to the predicted N abundance of $(C+N)_0$ of 8.12 dex. Considering this as within the uncertainties, the C, N and C-isotopic ratios suggest that the surface composition is as expected for the first dredge-up. Along with these quantities, the enrichment is also observed for He as well as Li. This high enrichment in Li, as mentioned above, would require the fresh synthesis along with the preservation of the pristine Li with which the star is formed. The initial/primordial Li-abundance is almost constant, i.e. $\log \epsilon$(Li)=3.3 dex, for the metallicity ranging from $-0.5<$[Fe/H]$<+0.2$ dex (Lambert & Reddy 2004; Randich et al. 2020). The main-sequence Li-abundance is expected to be diluted up to $\approx$1.9 to 1.6 dex for the mass range of 1M$\odot$ to 1.5M$\odot$ by the time they arrive at the base of the RGB phase. If the original Li-abundance with which the star is formed is less than the primordial value, then the abundance of Li during the RGB phase would further decrease. KIC 2305930 is also a binary star (Jorissen et al. 2020) which might have contributed by its binary companion in the enhancement of He and Li during mass-loss and mass transfer. Like in other program stars, La does not show any enhancement, indicating that the companion is not an AGB star or that it has not gone through the AGB phase.

HD 107484, by its position on the HR diagram, is probably at the tip of the RGB. From Table 11, the observed N abundance is about 8.1 dex that is close to the N abundance predicted for $(C+N)_0$ of 8.4 dex, within the uncertainties. This indicates that the surface composition is as expected for the first dredge up with the enhancement in He as well as Li (Table 6). Based on the criteria suggested by de la Reza (2025), the derived $\log g$ of about 2.5 cm s$^{-2}$ for HD 107484 is like an RC

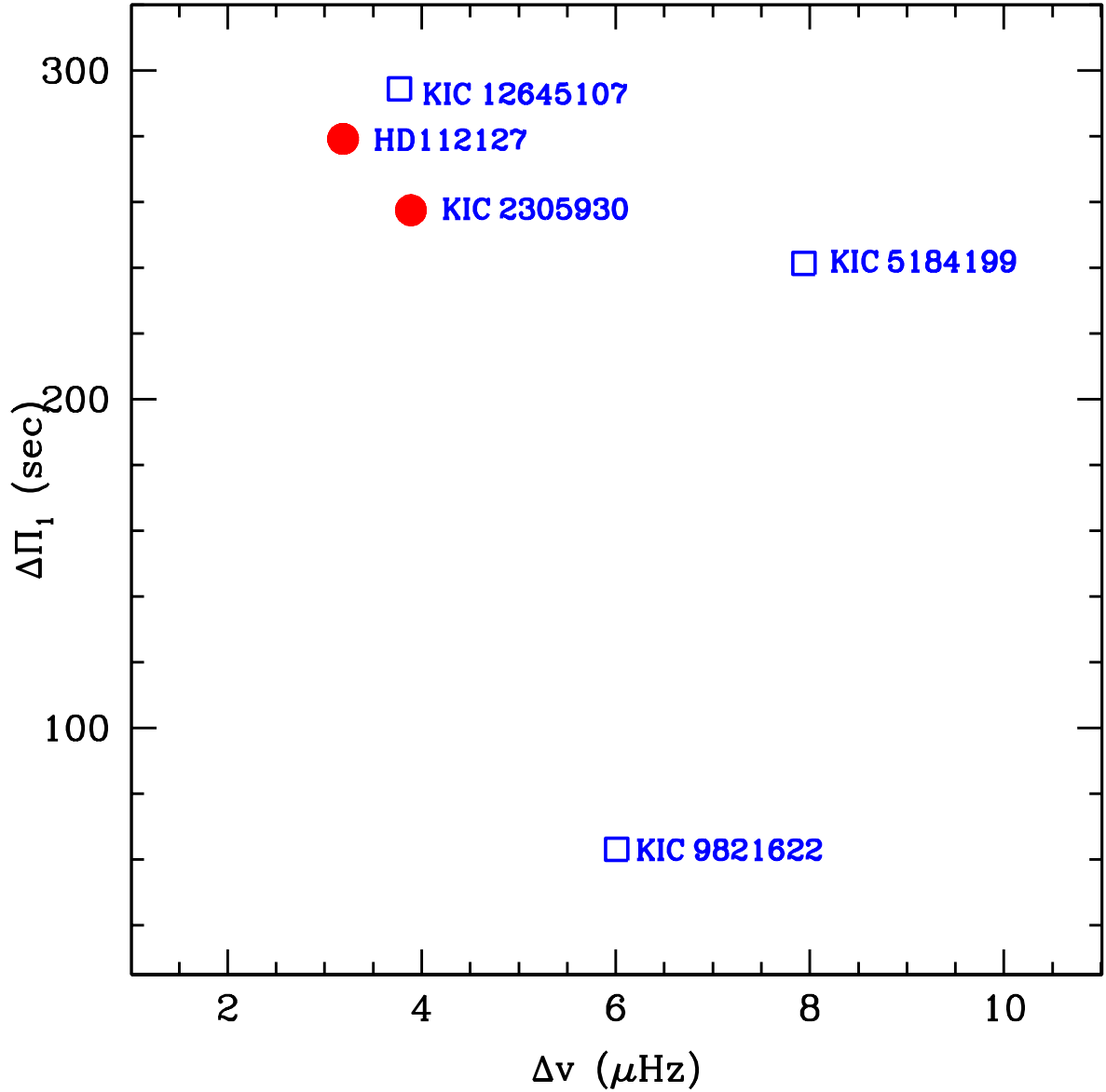

Fig. 12.— Figure shows the plot of $\Delta\Pi_1 - \Delta\nu$ for the programs stars with known evolutionary states using asteroseismology. The open blue squares are the giants with normal He/H ratio (0.1) and the filled red circles are the giants with He-enhancement.

star. The La is found to be moderately enhanced, but no information on the binarity is available.

HD 112127, is an RC star with He-enrichment of about He/H=0.4 for [Fe/H]$_{He}$ = −0.4 and log $\epsilon$(Li)=3.4±0.1 dex. The abundance ratios are [C/Fe]=0.30, and [N/Fe]=−0.05 and the $^{12}C/^{13}C$ ratio=22, as expected. The observed N-abundance is about 7.85 and this is very different from that predicted for CN- and ON-cycled N-abundance, which are 8.57 and 8.97 dex, respectively (see Table 11). However, the $^{12}C/^{13}C$ ratio is less than that observed in the MS stars, suggesting the surface enrichment with the dredge-up material. This would involve both preserving and fresh synthesis of Li, that is brought to the surface along with He and $^{13}C$. The enhancement in La is the same as that observed in HD 107484, with no information on binarity.

HD 77361, the He-enhancement is about 0.4 for [Fe/H]$_{He}$ = −0.3 dex, with log $\epsilon$(Li)=3.90 dex. The derived N-abundance, log $\epsilon$(N)=7.4 dex is very different from that predicted by CN and ON-cycle; 8.5 and 8.9 dex, respectively (see Table 11). The Li abundance of HD 77361 is greater than that expected in the MS phase. This requires the dredge-up of freshly synthesized Li. The derived $^{12}C/^{13}C$ ratio is about 8, which is close to the equilibrium value of 3.4 and as expected. The C and N are not the absolute indicators of mixing/dredge up unlike the $^{12}C/^{13}C$ ratio along with He- and Li-enhancement. The effective mixing observed could be due to the processes that take place in the bump phase of the RGB stars. The La is highly enhanced, indicating that the atmosphere is polluted by an AGB companion. But no information on the binarity is available for the program star. According to the criteria of de la Reza (2025), the log $g$ does not fall in both RGB or RC-group.

HD 30834 is an early-AGB star with He-enhancement of about 0.3 for [Fe/H]$_{He}$ = −0.45 dex and log $\epsilon$ (Li)=2.30 dex. The abundance ratios are [C/Fe]=−0.07, [N/Fe]=−0.6, [O/Fe]=0.04 and the $^{12}C/^{13}C$ ratio=13. The derived N-abundance, log$\epsilon$(N)=7.05 dex, is very different from that predicted by CN- and ON-cycle, respectively (see Table 11). The depletion in nitrogen is not as expected for the red giants but this would be the effect of either conversion of N to Ne on AGB (Karakas & Lattanzio 2003) or N may be less in the material with which the star is formed. Hence, the Li-, He- and the $^{13}C$ enrichment suggest mixing. The mass of HD 30834 is relatively higher (M=4.5M$_\odot$) compared to other program stars. Once the He in the core has been exhausted, the star's ascent on the AGB begins. On the AGB phase, the convective envelope deepens reaching the maximum depth, connecting the He-rich layer. This would trigger the production of Li by Cameron & Fowler process as discussed in (Charbonnel & Balachandran 2000). However, the observed Li abundance and the $^{12}C/^{13}C$ ratio are similar to other program stars as a thin hydrogen shell burning would only commence by the end of the early AGB phase followed by the thermally pulsating phases. Hence, the derived abundances in the early-AGB stars are similar to that observed in the RGB stars. The [La/Fe] shows enhancement which is as expected for an AGB star.

HD 181475 is a supergiant star enhanced in He. This star is He-enriched but not Li-rich unlike the other He-enriched program stars discussed above. The Li-abundance is about log $\epsilon$ (Li)=−0.8 dex.

The abundance ratios are [C/Fe]=0.16, and [N/Fe]=−0.54. The $^{12}$C/$^{13}$C ratio couldn't be derived for the star as the spectra did not cover the CN-band at about 8000Å. From Table 11, the observed N-abundance is about 7.4 and this is very different from that predicted for (C+N)$_0$=8.6 and (C+N+O)$_0$=9.0 dex, respectively. As this is a supergiant, there would be multiple episodes of convection and mixing happening, which would result in the He-enhancement and Li-depletion.

Discussed above are the He- as well as Li-enhanced giants, but, we also observe stars that are Li-rich but not He-rich. The majority of our sample stars have a range in Li-abundance; no-Li, Li-rich and super Li-rich, with no He-enhancement. The program stars that are having no-Li, with log$\epsilon$(Li)≤1.00 dex are; KIC 5184199, Ups02 Cas, HR 4813, and $\alpha$ Boo. The stars that have log$\epsilon$(Li)≥1.18 dex are grouped as Li-rich, they are HR 6766, KIC 9821622 HD 40827, HR 334, HD 205349. And the giants with log$\epsilon$(Li)≥2.5 dex are grouped as super Li-rich, they are: KIC 12645107, HD 19745, HD 214995, HDE 233517, and HD 39853. The discussion of these three groups are as follows.

### 7.1.2. Giants with no/less Li

Our sample giants; KIC 5184199, Ups02 Cas, HR 4813 and Arcturus/$\alpha$ Boo has no Li, having the log$\epsilon$(Li)≤1.00 dex and $^{12}$C/$^{13}$C < 25, as expected. Though these stars have different [C/Fe] and [N/Fe] abundance ratios indicating different scales of mixing, their Li and $^{12}$C/$^{13}$C ratios are as expected for the first dredge up in RGB stars. Hence, $^{12}$C/$^{13}$C is a better indicator of mixing. For these stars, the derived N-abundance is not as predicted for the CN-cycling (see Table 11).

### 7.1.3. Li-rich Giants

Another subgroup is with Li-rich abundance $1.18 < \log \epsilon$(Li) $< 2.5$. These are; HR 6766, KIC 9821622, HD 40827, HR 334, HD 205349. Among these giants, HR 6766, KIC 9821622, and HD 40827 show efficient mixing with greatly enhanced N and depleted C, unlike the other two stars, HR 334 and HD 205347 (see Table 11). Surprisingly, despite high enhancement in N and highly depleted C in the weak G-band star, HR 6766, the Li is measurable and like that of an MS star. This is a puzzle, this Li preserved or freshly synthesized? The observed abundances in KIC 9821622 and HD 40827 pose the same problem as well. The other two giants, HR 334 and HD 205349 neither show efficient mixing nor Li-depletion completely (see Table 11) – indicating that Li here is preserved from the MS phase.

### 7.1.4.  Super Li-rich giants

This a group of giants with super Li-rich abundances with no He-enhancement, and these are; KIC 12645107, HD 19745, HD 214995, HDE 233517 and HD 39853. KIC 12645107 is an RC star with [C/Fe]=−0.45 and [N/Fe]=0.57. The $^{12}$C/$^{13}$C ratio is about 6, with $\log \epsilon$(Li)=3.72±0.1 dex. The observed N-abundance ($\log \epsilon$(N)=8.3 dex) is similar to that predicted for the CN-cycled abundance ($\log \epsilon$(N)=8.42 dex). These abundances are as expected for giants except for Li, which is similar to the primordial value. For other giants of this sample, the observed N-abundance is similar to that predicted for CN-cycling (refer to Table 11) as expected, except for HD 19745 and HD 39853. The Super Li-rich abundance is probably the result of preservation from the pristine material or/and the fresh synthesis in these stars.

## 7.2.  Discussions

As discussed above, our sample stars belong to different stages spanning from the base of the RGB to the early-AGB phase. The metallicity of the sample is $−0.6 < $[Fe/H]$ < +0.2$ dex. The main aim of this study includes the investigation of photospheric He-enhancement in principal, along with the CNO-abundances and the $^{12}$C/$^{13}$C ratio, in tandem with their Li-abundances. The main discovery of this work is that all the He-enhanced red giants including an early-AGB star are super Li-rich, except for the supergiant star. This indicates that the photospheric He-enrichment is accompanied by the Li-enrichment in the giants, as we envisage. Our Li-rich sample stars, evince that there exists two or more processes, at least, responsible for Li-enrichment in giants, one to enrich only Li, and another to enrich He along with Li.

Mallick et al. (2025) study reports the correlation between the high photospheric Li-abundances and the strong chromospheric He I $\lambda$10830Å absorption line strengths in Kepler field giant stars. The correlation of high photospheric Li-abundance with the strong chromospheric He I line is found exclusively among RC giants. Our results show that the high photospheric Li-abundances are in correlation with the photospheric He-enrichment. We looked for the equivalent width of the He I line for our program stars in their sample, and found two stars in common with their study. These are Li-rich RC stars, KIC 2305930 and KIC 12645107. In our study we find that one, KIC 12645107, is a He-normal and another, KIC 2305930, is He-enhanced. The He-enhanced KIC 2305930 is having a strong He I line $\lambda\lambda$ 10830Å of EW$_{He}$=300±10 mÅ with $\log \epsilon$(Li)=4.03 dex, while the He-normal KIC 12645107 is having a moderately strong He I line at $\lambda\lambda$ 10830Å of EW$_{He}$=95±8 mÅ with $\log \epsilon$(Li)=3.72 dex. For these two stars, relatively, it is observed that the chromospheric $\lambda\lambda$ 10830Å He I line is more pronounced in the He-enriched star than in the He-normal one. This is an indication that the chromospheric He-activity in RC stars is correlated with the photospheric He-enhancement with confirmed super Li-rich abundances.

However, as observed from our results, He-enhancement is not only confined to RC stars but also observed in RGB stars. Our five He-rich RGBs, include two RC stars, one from the RGB tip and another from the RGB-bump phase along with an early-AGB star. Since the lifetimes of the RC phase are two to five times longer than the RGB phase (Iben 1967), it is likely to find more RC stars with only Li-enrichment as well as both He- and Li-enrichment. Further studies on a significant sample including both the RC and RGB samples with known evolutionary states would help to discern the scenarios responsible for these observations.

The derived CN abundances of the program stars show no uniformity. The relation of $\log \epsilon(C)$ and $\log \epsilon(N)$ versus [Fe/H] for the program stars are shown in the bottom panel of Figures 13 and 14, respectively. However, the $^{12}C/^{13}C$ ratios derived for the sample giants, that is $\leq 22$, is as expected for giants. The low-mass sample giants with the M$\sim$M$\odot$ are expected to operate only the pp-chain in their interiors during their main-sequence phase. Hence, the processed material, especially CNO, that is brought to the surface during the dredge-up is almost similar to the primordial/main-sequence composition. This is as observed in some of our program stars. A few program stars are intermediate-mass stars that operate CN-cycle along with the pp-chain at the center during the main-sequence phase. This material that is enriched with CN-cycled material is dredged-up and mixed with the surface (Clayton 1968; Choudhuri 2010).

Out of six He-enriched giants of our study, two giants show enhancement in N as expected for CN-cycle. However, in the top panel of Figure 15, with [C/N] vs. $\log \epsilon(Li)$ relation, the [C/N] ratio shows a nearly constant trend that is less than the solar value, suggesting C-depletion and N-enhancement relative to Fe, with respect to solar. The He-rich stars are in the Super Li-rich regime in the Luminosity versus Li relation as shown in the bottom panel of Figure 15. The other elemental abundances derived for the program stars are as expected for the giants with about nine program stars showing enhancement in La from 5 to 10 times solar. The He-enhanced giants, HD 77361 and HD 30834 show enhancement in La abundance with [La/Fe] $\sim$ 1.0 dex. HD 112127 and HD 107484 show only moderate enhancements in [La/Fe]. The [La/Fe] is like solar for KIC 2305930, and the supergiant star shows a very high enhancement of about [La/Fe]=1.45 dex, as expected for an AGB star. Out of the He-normal giants, about 10 giants show moderate enhancement and other four of them show no enhancement in [La/Fe]. Since our La abundance is from a single line, we do not draw any conclusions based on this measurement.

Multiple scenarios are proposed to explain the Li-enrichment in giants. First, preserving the Li from the pristine (if different from primordial) material with which the star is formed. Second, the fresh production of Li during the main-sequence and RGB that is brought to the surface through the convection and dredge up. The third efficient channel for Li-enrichment is via planet engulfment.

As the star evolves off the main-sequence phase and moves along the giant branch its core shrinks and the outer envelope expands engulfing its nearest planets. For example, the Earth has Li that is one to two orders greater than that observed in the Sun (Alexander 1967). If a giant engulfs an earth like planet, the host star becomes Li-rich. In the atmosphere of a Jupiter like planet, the Li is diluted and the helium is enhanced ((He/H2)=0.16±0.03 (Ben-Jaffel & Herbert 2010)). Such planets do not change the Li-abundance in the atmosphere of the host star on engulfment (Alexander 1967). However, for a giant to become Li-rich on engulfing a planet, the mass of the giant/host star as well as the mass of the planet are crucial.

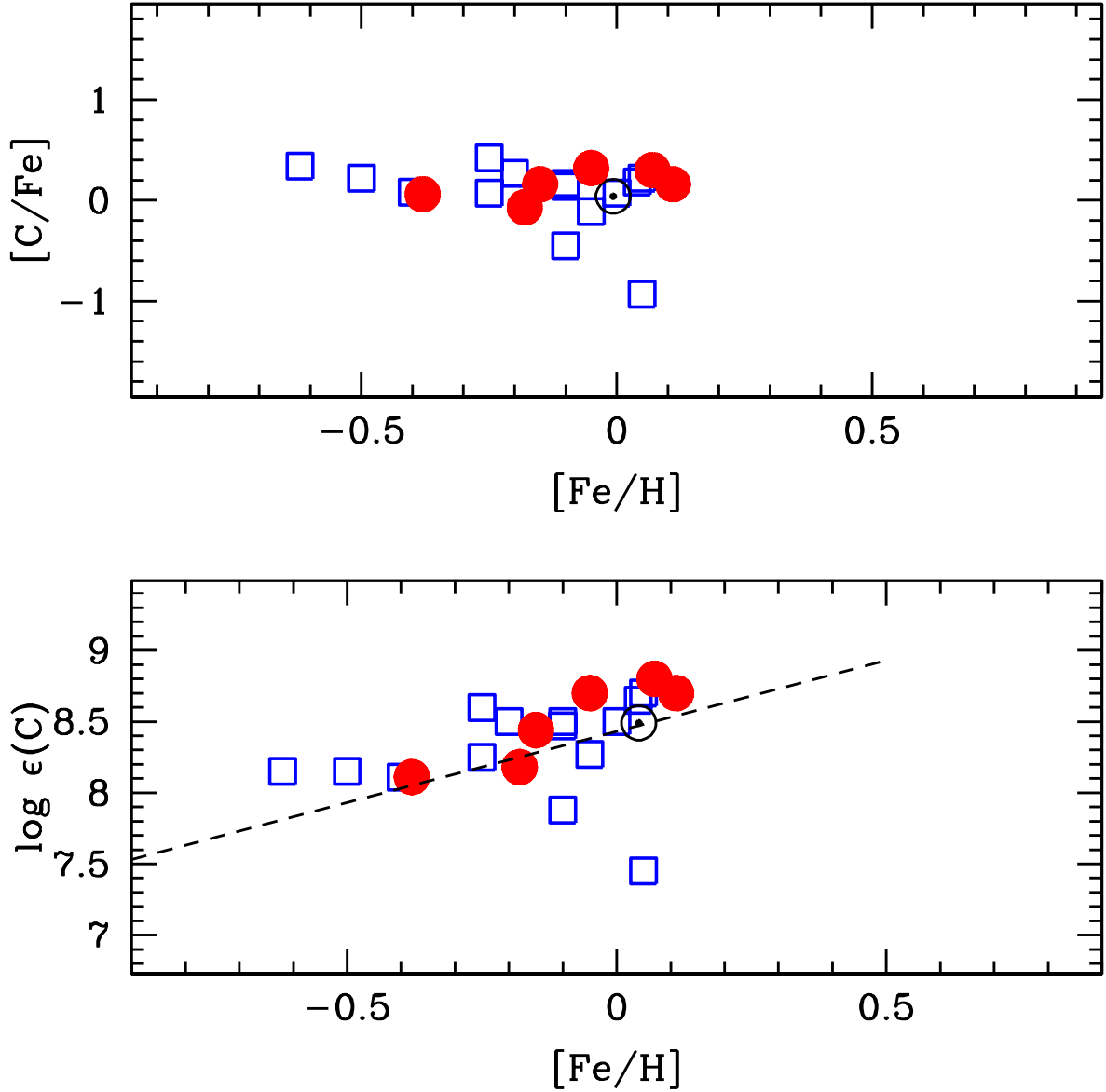

Fig. 13.— Bottom Panel: Figure shows the abundance of carbon derived for the program stars versus [Fe/H]. The open blue squares are the giants with normal He/H ratio (0.1) and the filled red circles are the giants with He-enhancement. The dashed line is for the initial carbon with respect to metallicity. And ⊙ represents the Sun. Top Panel: Figure shows the C-abundance relative to Fe, w.r.t. Sun, [C/Fe] for the program stars versus [Fe/H]. The open blue squares are the giants with normal He/H ratio (0.1) and the filled red circles are the giants with He-enhancement. And ⊙ represents the Sun.

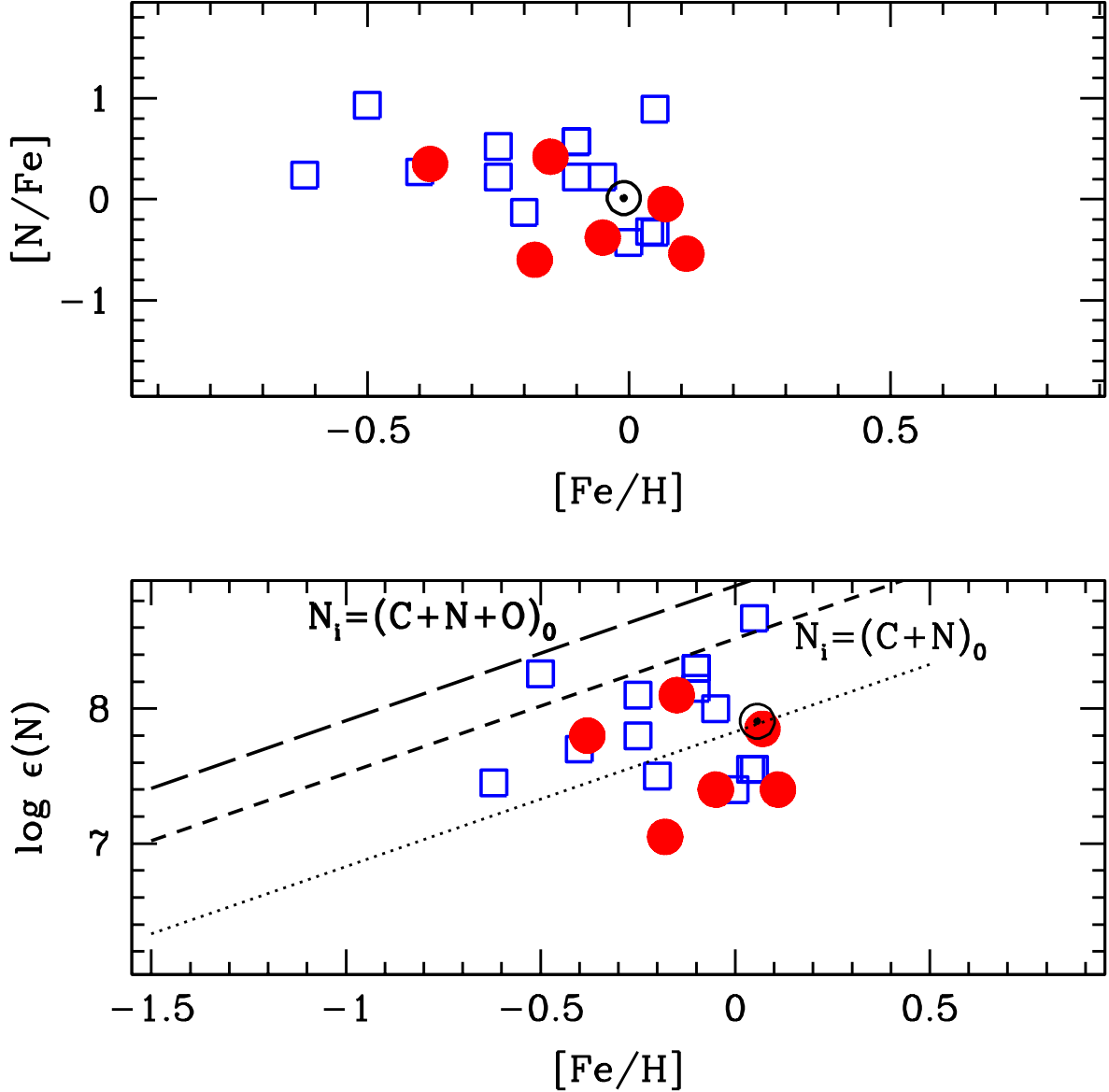

Fig. 14.— Bottom Panel: Figure shows the abundance of Nitrogen ($\log \epsilon(N)$) derived for the program stars versus [Fe/H]. The open blue squares are the giants with normal He/H ratio (0.1). The filled red circles are the giants with He-enhancement. The short dashed line for $\log\epsilon(N)_i=(C+N)_0$, and the long dashed line for $\log\epsilon(N)_i=(C+N+O)_0$, where $\log\epsilon(N)_i$ is for CN- and ON-cycles for initial CNO-abundances, with [Fe/H] the metallicity as denoted. Top Panel: Figure shows the abundance ratio of Nitrogen, [N/Fe] derived for the program stars versus [Fe/H]. The open blue squares are the giants with normal He/H ratio (0.1) and filled red circles are the giants with He-enhancement. And $\odot$ represents the Sun.

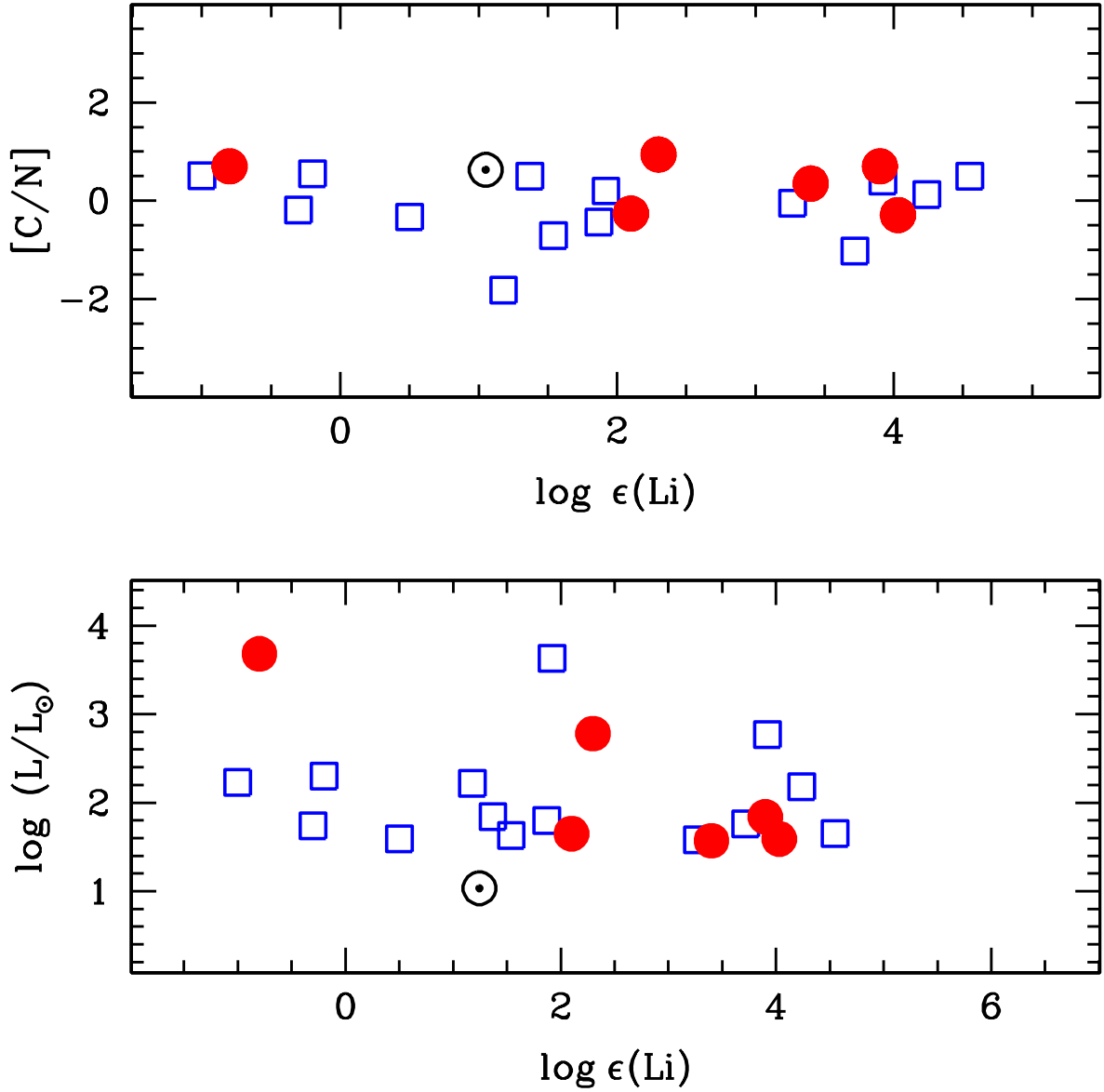

Fig. 15.— Bottom Panel: Figure shows the Luminosity (L/L⊙) versus the Li-abundance log ε(Li) for the program stars. The blue open squares are the stars with normal He/H ratios and the filled red circles are the He-enhanced ones. Top Panel: Figure shows the ratio of C and N w r.t. Sun [C/N] versus the Li-abundance for the program stars. The open blue squares are the stars with normal He/H ratio (=0.1) and the filled red circles are the He-enhanced ones.

In this work, the most probable scenario that deemed apt for the five He- as well as Li-enriched giants would be the fresh production of Li in the interiors of the RGB stars that is swiftly dredged up to the surface along with He from the deeper layers (Pilachowski et al. 1990).

To explore these plausible scenarios how the Li survives from the hotter temperatures needs to be addressed. Also, during the Cameron-Fowler reaction, we need to explore how the $^7$Be is transported to cooler regions such that the $^7$Li produced by the decay of $^7$Be can survive (Charbonnel & Balachandran 2000; Gao et al. 2022). In this study, it is observed that all the He-enhanced red giants are super Li-rich, but all the Li-rich/super Li-rich giants are not He-rich. Kumar et al. (2020), in their study of a large sample of giants, suggest that all the low-mass stars undergo a lithium production phase between the tip of the RGB and the RC phase (see their Figure 3). Their findings suggest a new limit for Li-enrichment in red-clump stars (the evolutionary states are determined from asteroseismology), that is $\log \epsilon(\text{Li}) > -0.9$ dex. However, as reported in many studies (mentioned above), including this work, giants show significant Li-enrichment all across the giant branch for which the evolutionary states are may or may not be determined solely from asteroseismology. The notable derived Li-abundances among the sample stars of known evolutionary states from asteroseismology are about; $<0.50$ dex for KIC 5184199 (RC-star), 3.72 dex for KIC 12645107 (RC-star), 4.03 dex for KIC 2305930 (RC-star), and 3.4 dex for HD 112127 (RC-star), and 1.54 dex for KIC 9821622 the bump star. The RGB-bump star, KIC 9821622 shows the Li-enrichment which is also reported by Jofré et al. (2015). This indicates that the giants with Li-enrichment are observed in the RGB-bump, -clump, and as well as in other stages of RGB-evolution.

## 8. Concluding Remarks

The primary objective of this work is to detect and measure the He-enhancement in the photospheres of cool giants, and further investigate any connection of He-enhancement with Li-enrichment. The program stars comprise a random sample of giants; 16 RGB, 2 early-AGB, and 2 Supergiants, with and without Li-enrichment. With the help of the observed optical high-resolution spectra of the program stars, a-detailed abundance analysis was carried out to determine the abundances of the key elements; He, Li, C, N, O, among many others including the C-isotopic ratios. Using the difference in the Mg-abundances derived from the Mg I and the MgH lines, the He-enhancement is explored. These are the best features to measure He-abundance, than the C I lines and CH-band, for giants. Out of these sample giants, we identified about 4 red giants, an early-AGB star and a supergiant star to be He-enriched. Surprisingly, all these He-enhanced red giants along with the early-AGB star are super Li-rich as well, except for the supergiant. This is a clear suggestion that the He-enhancement is accompanied by a significant amount of Li-enrichment/production in red giants including early-AGB stars. The photospheric He-enhancements in these stars also appear to be associated with their chromospheric He-activity. Our results show that the Li-enrichment is not necessarily associated with He-enhancement. This again suggests that there must be multiple channels that make a star Li-rich. Compared to the giants, supergiants undergo structural changes that affect the surface chemical composition. Hence, observing no-Li in the atmosphere of the supergiants is as expected despite the He-enhancement. This work strongly evinces the process of fresh synthesis of Li and dredging up to the surface, in the giant phase for the observed Li-enrichment as one of the channels. The depth of the convection zone would be the deciding factor for He-enhancement during the dredge-up. These sound observational results await more thorough stellar interior models.

Table 11.   Observed and the predicted CNO abundances for the program stars. $\log \epsilon(C+N)_0$ and $\log \epsilon(C+N+O)_0$ are the predicted N-abundances for the CN- and ON-cycles from the stars' initial CNO abundances.

| Stars | [Fe/H] | $\log \epsilon(O)$ | $\Delta^{a}$ | $\log \epsilon(N)$ | $\log \epsilon(C)$ | $\log \epsilon(N)_{predicted}$ | | $[(C+N+O)/Fe]_{obs}$ | $\log \epsilon(Li)$ | He/H |
|---|---|---|---|---|---|---|---|---|---|---|
| | | | | | | $\log \epsilon(C+N)_0$ | $\log \epsilon(C+N+O)_0$ | | | |
| Sun | 0.0 | 8.69 | $\cdots$ | 7.83 | 8.43 | $\cdots$ | $\cdots$ | 8.91 | 1.05 | 0.1 |
| KIC 5184199 | -0.05 | 8.87 | 0.18 | 8.0 | 8.27 | 8.47 | 8.86 | 0.15 | <0.50 | 0.1 |
| KIC 12645107 | -0.10 | 9.0 | 0.32 | 8.3 | 7.88 | 8.42 | 8.81 | 0.29 | 3.72 | 0.1 |
| HR 6766 | −0.05 | 8.73 | 0.09 | 8.67 | 7.45 | 8.45 | 8.85 | 0.14 | 1.18 | 0.1 |
| KIC 9821622 | −0.5 | 8.84 | 0.65 | 8.26 | 8.15 | 8.00 | 8.40 | 0.60 | 1.54 | 0.1 |
| Ups02 Cas | -0.4 | 8.72 | 0.03 | 7.7 | 8.11 | 8.12 | 8.51 | 0.46 | <-0.3 | 0.1 |
| KIC 2305930 | -0.40 | 9.5 | 0.81 | 7.8 | 8.11 | 8.12 | 8.51 | 1.01 | 4.03 | 0.4 |
| HD 19745 | 0.0 | 8.60 | −0.13 | 7.4 | 8.5 | 8.50 | 8.90 | 0.0 | 4.55 | 0.1 |
| HD 40827 | -0.1 | 9.07 | 0.38 | 8.30 | 8.47 | 8.40 | 8.80 | 0.42 | 1.87 | 0.1 |
| HD 214995 | -0.1 | 8.95 | 0.36 | 8.15 | 8.50 | 8.40 | 8.80 | 0.31 | 3.27 | 0.1 |
| HDE 233517 | -0.25 | 9.34 | 0.65 | 8.1 | 8.25 | 8.25 | 8.65 | 0.81 | 4.24 | 0.1 |
| HD 107484 | -0.15 | 9.04 | 0.35 | 8.1 | 8.44 | 8.40 | 8.76 | 0.141 | 2.10 | 0.4 |
| HR 334 | 0.04 | 8.85 | 0.12 | 7.55 | 8.65 | 8.54 | 8.94 | 0.31 | 1.37 | 0.1 |
| HR 4813 | 0.04 | 8.84 | 0.13 | 7.55 | 8.70 | 8.54 | 8.94 | 0.38 | < −0.2 | 0.1 |
| HD 112127 | 0.07 | 8.95 | 0.21 | 7.85 | 8.8 | 8.57 | 8.97 | 0.23 | 3.4 | 0.4 |
| $\alpha$ Boo | -0.6 | 8.75 | 0.06 | 7.45 | 8.15 | 7.92 | 8.31 | 0.58 | < −1.0 | 0.1 |
| HD 205349 | −0.25 | 8.85 | 0.41 | 7.80 | 8.6 | 8.25 | 8.65 | 0.30 | 1.92 | 0.1 |
| HD 77361 | 0.0 | 8.75 | 0.06 | 7.4 | 8.7 | 8.50 | 8.90 | 0.21 | 3.90 | 0.4 |
| HD 30834 | −0.18 | 8.5 | −0.03 | 7.05 | 8.18 | 8.32 | 8.72 | 1.36 | 2.3 | 0.3 |
| HD 181475 | 0.1 | 9.1 | 0.31 | 7.40 | 8.7 | 8.60 | 9.00 | 0.11 | < −0.80 | 0.5 |
| HD 39853 | −0.20 | 8.53 | 0.0 | 7.5 | 8.5 | 8.32 | 8.71 | 1.47 | 3.90 | 0.1 |

$^{a}\Delta = \log(\epsilon(O)_{obs}/\epsilon(O)_0)$.

## 9. Acknowledgments

B.P.H. acknowledges and thanks the Women Scientist Scheme-A (WOS-A), Department of Science and Technology (DST), India, for support through the grant DST/WOS-A/PM-1/2020, and the Indian Institute of Astrophysics for support through the Visiting Scientist Program. GP thanks DST/SERB for their support through CRG grant CRG/2021/000108. We are grateful to the anonymous referee for a detailed and constructive report that has improved the presentation of the results considerably. This research is based on the following sources: high-resolution spectral data obtained with the 2 m Himalayan Chandra Telescope (HCT); the facilities at IAO and CREST are operated by the Indian Institute of Astrophysics, Bengaluru. Based on the spectral data retrieved from the ELODIE archive at Observatoire de Haute-Provence (OHP, http://atlas.obs-hp.fr/elodie/). Observations collected at the European Organisation for Astronomical Research in the Southern Hemisphere under ESO programmes, 096.D-0421(A), 089.D-0189(A), 098.A-9029(A). This research uses the spectral data from the Nicholas U. Mayall 4-meter Telescope at Kitt Peak National Observatory (KPNO) The authors are very thankful to Dr. Bharat Kumar Yerra and Prof. Eswar Reddy for providing them with the spectrum of HD 77361, obtained from the McDonald Observatory. We thank Prof. Bertrand Pluz for kindly providing us with the CH line list. We thank Ms. Anohita Mallick for helping us to derive the asteroseismic parameters from TESS data for the program stars. We thank Prof. Kurucz and Prof. Carlos Allende Prieto for helping us in obtaining the He-enhanced models.

## Appendix Material

## A.  CNO Abundances for the program stars

To understand the stellar evolution of the giants, the abundances of the key elements: carbon, nitrogen and oxygen is very crucial. For all our program stars the carbon, nitrogen and oxygen abundances are derived using both the atomic and molecular lines.

**Carbon:** For deriving the carbon abundance, the permitted and the forbidden lines were used.

The permitted lines used are: $\lambda\lambda$4775.91, 5052.17, 5380.34, 6587.60, 8335.15, 9603.03, 9850.34Å. And,

the forbidden [C I] line is: $\lambda$8727.12Å. These C I lines are from the Lambert (1978); Wiese et al. (1996); Abia et al. (2012); Luck & Lambert (1981); Lambert & Ries (1981). These studies have used these lines to determine the CNO abundances for Sun and the giants. Though, the C I lines in the spectra of the program stars are weak, but they are well represented. For all these lines, the equivalent widths were measured in the spectrum of Arcturus and the lines were inverted for deriving the log $gf$ values knowing the accurate carbon abundance (Ramírez & Allende Prieto 2011; Hema et al. 2018). The derived log $gf$ values are in good agreement with those given in

Wiese et al. (1996).

For all the program stars, minimum of two C I and maximum of eight lines (as listed above) were found, subjected to wavelength span of the spectrum. The C I that are very weak/within the noise level of the continuum were discarded. However, more weight is given to C I: λ5380.34Å and [C I]: λ8727.12Å lines, as these are less blended and relatively stronger/well represented than other spectral lines of carbon. The carbon abundances derived in this study are in good agreement with those available in the literature, including for the Arcturus.

**Oxygen:** For the determination of oxygen abundance in the program stars the spectral lines used are:

The permitted lines: $\lambda\lambda$6158.18, 9265.90, 7771.94, 7774.17, 7775.39Å.

The forbidden lines: $\lambda\lambda$5577.34, 6300.30, 6363.77Å.

These lines are also from Lambert (1978), for which the $\log gf$ values are verified from the Solar and Arcturus spectrum. The derived $\log gf$ values are in good agreement with those from Wiese et al. (1996). The oxygen abundance derived in this study for Arcturus are in excellent agreement with those derived by Ramírez & Allende Prieto (2011) within the uncertainties. In many of our program stars spectra, minimum of three O I and two [O I] lines were present allowing us to make a fair measurement of oxygen abundance.

**Nitrogen:** Several N I lines from the NIST data base were compiled in the spectral range of the program stars. The N I lines have relatively higher LEP and are not expected to be seen in the cooler giants. The N I lines were not present in many of our program stars, but for a few giants with relatively warmer $T_{\text{eff}}$ and significant enrichment in nitrogen. However, the best way to determine the nitrogen abundance is from the CN bands. For many of our program stars, the nitrogen abundance is derived from CN-red and -violet bands as described below. Hence, using the program star's derived carbon abundance from C I lines, the nitrogen abundance was measured from the CN band.

**CN-band:** The CN-red and -violet bands provide many weak and unblended lines enabling the determination of nitrogen abundance provided with an independent determination of the carbon abundance. The carbon abundances for the program stars are independently and accurately derived from the C I and [C I] lines. The nitrogen abundance has been mainly determined from the CN-violet system in the wavelength window: 4212 to 4217Å and the $^{12}C/^{13}C$ ratio is determined from the CN-red system in the wavelength window: λ8000 to 8010Å. This spectral window covers both the $^{12}C^{14}N$ as well as $^{13}C^{14}N$ lines. The line list for both red and violet bands were adopted from Sneden et al. (2014). By using the appropriate **ATLAS9 (Kurucz 1998)** model atmospheres with the MOOG *synth* code, the synthesized spectra were matched with the observed spectra of program stars by varying the nitrogen abundance, and keeping the carbon abundance from the C I lines fixed. This data and procedure are validated by synthesizing the Arcturus spectrum. This derived nitrogen abundance from CN bands for Arcturus is in good agreement with that reported by Sneden et al. (2014) within the uncertainties (see the bottom panel of Figure 3 therein).

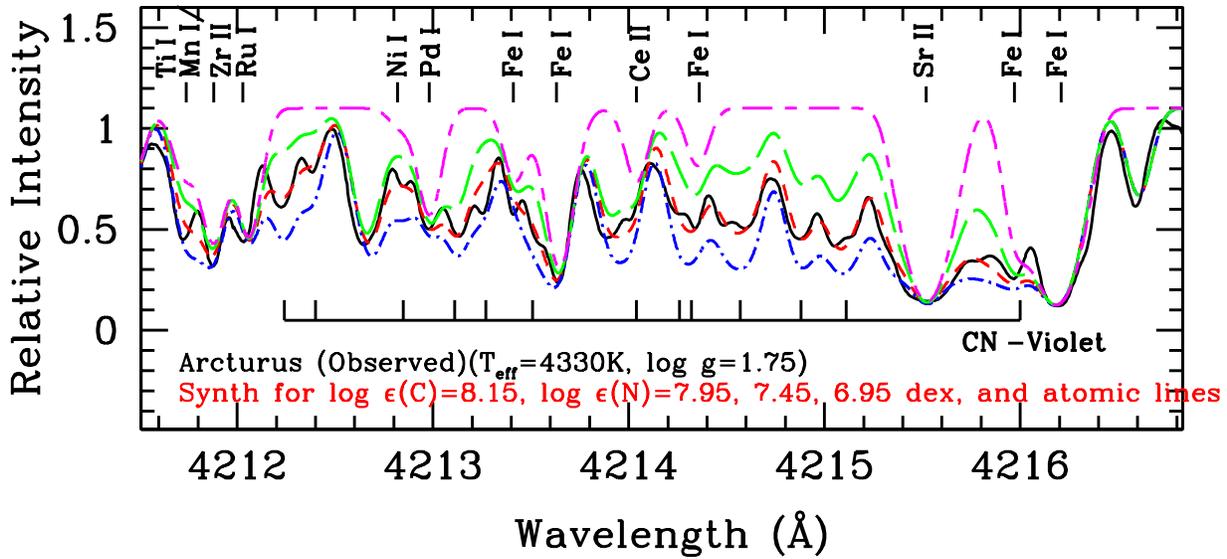

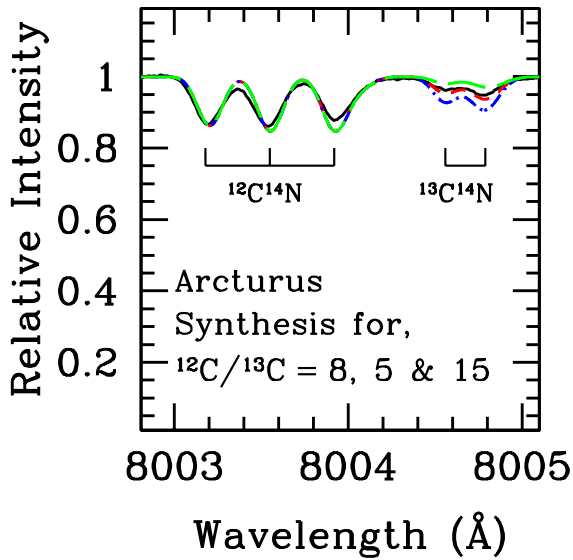

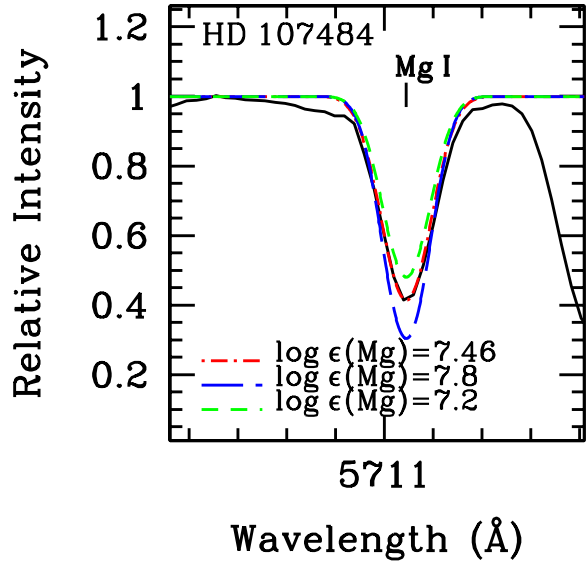

Fig. A.1.— **Top panel:** Figure show the syntheses of the CN band for Arcturus. The synthesis shown with short dashed red line is for the best fit N-abundance. The syntheses are also shown for the other two values of N-abundance for comparison, with long dashed green line and dash-dotted blue line. The long-short dashed magenta line is for pure atomic lines. **Bottom left:** Figure shows the determination of the $^{12}C/^{13}C$ ratio from $^{12}C^{14}N$ and $^{13}C^{14}N$, lines for Arcturus. The best fit is shown with the dashed red line for the $^{12}C/^{13}C=8$, the synthesis for other two ratios are shown for comparison. **Bottom right:** Figure shows the synthesis of the Mg I line, $\lambda\lambda5711$Å for the program star HD 107484, as an example.

Table A.1.   The list of newly included atomic lines along with those adopted by Hema et al. (2018). Table gives the species, wavelength ($\lambda$), lower excitation potential ($\chi$) and the transition probabilities ($\log gf$).

| Species | $\lambda(\text{Å})$ | $\chi(\text{eV})$ | $\log gf$ |
|---------|---------|---------|---------|
| Li I | 6103.540 | 1.840 | +0.106 |
| Li I | 6707.800 | 0.000 | +0.020 |
| C I | 8335.150 | 7.680 | −0.440 |
| C I | 9078.280 | 7.480 | −0.570 |
| C I | 9111.800 | 7.490 | −0.300 |
| N I | 7442.290 | 10.330 | −0.330 |
| N I | 7468.310 | 10.340 | −0.160 |
| N I | 8594.010 | 10.680 | −0.320 |
| N I | 8683.400 | 10.330 | +0.110 |
| N I | 8703.250 | 10.330 | −0.290 |
| N I | 8718.830 | 10.340 | −0.260 |
| O I | 7771.940 | 9.150 | +0.330 |
| O I | 7774.170 | 9.150 | +0.190 |
| O I | 7775.390 | 9.150 | −0.030 |
| Fe I | 4635.850 | 2.845 | −2.360 |
| Fe I | 4690.140 | 3.686 | −1.640 |
| Fe I | 4802.880 | 3.642 | −1.510 |
| Fe I | 5295.310 | 4.420 | −1.590 |
| Fe I | 5329.990 | 4.070 | −1.220 |
| Fe I | 5403.820 | 4.076 | −1.030 |
| Fe II | 5414.070 | 3.221 | −3.480 |
| Fe I | 5466.990 | 3.573 | −2.23 |
| Fe I | 5483.100 | 4.154 | −1.410 |
| Fe I | 5525.540 | 4.230 | −1.080 |
| Fe I | 5661.348 | 4.280 | −1.750 |
| Fe I | 5705.460 | 4.300 | −1.360 |
| Fe I | 5741.850 | 4.256 | −1.670 |
| Fe I | 5778.458 | 2.590 | −3.450 |
| Fe I | 5784.661 | 3.400 | −2.530 |
| Fe I | 5809.220 | 3.884 | −1.610 |
| Fe I | 5849.69 | 3.695 | −2.930 |
| Fe I | 5855.090 | 4.608 | −1.480 |
| Fe I | 5856.100 | 4.294 | −1.560 |
| Fe I | 5858.790 | 4.220 | −2.180 |
| Fe I | 5859.600 | 4.550 | −0.610 |
| Fe I | 5862.370 | 4.550 | −0.250 |
| Fe I | 6027.060 | 4.070 | −1.170 |
| Fe I | 6082.710 | 2.230 | −3.550 |

Table A.1—Continued

| Species | $\lambda(\text{Å})$ | $\chi(\text{eV})$ | $\log gf$ |
|---------|---------|--------|-----------|
| Fe I | 6159.380 | 4.610 | −1.830 |
| Fe I | 6165.360 | 4.143 | −1.460 |
| Fe I | 6271.283 | 3.330 | −2.700 |
| Fe I | 6581.214 | 1.48 | −4.680 |
| Fe I | 6591.330 | 4.593 | −1.950 |
| Fe I | 6608.040 | 2.279 | −3.910 |
| Fe I | 6713.750 | 4.795 | −1.390 |
| Fe I | 6725.360 | 4.103 | −2.170 |
| Fe I | 6733.150 | 4.638 | −1.400 |
| Fe I | 6752.711 | 4.640 | −1.200 |
| Fe I | 6793.260 | 4.076 | −2.330 |
| Fe I | 6854.820 | 4.593 | −1.930 |
| Fe I | 6857.250 | 4.076 | −2.040 |
| Fe I | 7189.150 | 3.070 | −2.770 |
| Fe I | 7401.690 | 4.186 | −1.600 |
| Fe I | 7807.920 | 4.990 | −0.510 |
| Fe II | 5234.620 | 3.221 | −2.22 |
| Fe II | 5425.260 | 3.200 | −3.160 |
| Fe II | 6149.250 | 3.889 | −2.630 |
| Fe II | 6247.560 | 3.892 | −2.270 |
| Fe II | 6369.460 | 2.891 | −4.020 |
| Fe II | 6378.260 | 4.150 | −1.050 |
| Fe II | 6432.680 | 2.891 | −3.520 |
| Fe II | 6456.390 | 3.903 | −2.060 |
| Fe II | 6516.077 | 2.891 | −3.52 |
| Fe II | 7711.720 | 3.903 | −2.450 |